# Beam performance and instrumentation studies for the ProtoDUNE-DP experiment of CENF

N. Charitonidis[*, 1], Y. Karyotakis[2, 1] and I. Efthymiopoulos[1]

[1] CERN, 1211 Geneva 23, Switzerland
[2] LAPP, Lab. d'Annecy-le-Vieux de Physique de Particules, 74941 Annecy-le-Vieux CEDEX - France

\* Corresponding author : nikolaos.charitonidis@cern.ch

### Abstract

In this note, we address the beam performance (particle content, rates) with emphasis on the momentum determination and particle identification methods for the new **H2-VLE** (Very Low Energy) beam line that will serve the double phase ProtoDUNE experiment (also known as WA105), in the framework of the CENF project. The proposed instrumentation is configured to achieve an optimal pi/K/proton separation over the full spectrum of provided beam energies, from 0.4 GeV up to 12 GeV, as well as precise momentum measurement to a percent level, if required by the experiment. This note focuses on the H2-VLE beam line for the Double Phase ProtoDUNE experiment, however the same approach can be implemented for the H4-VLE beam, since the design of the two beam lines is very similar.

## 1. Beam Layout

The new beam line, designated '**H2-VLE**' (and **similarly 'H4-VLE'**) constitutes an extension of approximately ~ 40 meters of the existing infrastructure of H2 (H4) beam line of the North Area Complex of SPS, designed to provide low-energy particles in the range of sub-GeV to few GeV to the installed CENF experiments. In each case, the low energy particles are produced through the interaction of a secondary beam of high-intensity (~) and momentum (~60-80 GeV/c) with a target material. Typically, for producing a positive low-energy hadron beam, a secondary hadron beam of 80 GeV consisting of 70% pi[+], 25% p and 5% of K+ with ~1% $\Delta p/p$ momentum spread, and with intensity of few $10^6$ particles per pulse is used. Similarly, for producing low-energy electrons, a secondary, relatively pure electron beam of 50-60 GeV is used. In all cases, the secondary beam is transported through the existing H2 (and H4) beam lines on the secondary target. The **V**ery **L**ow **E**nergy Extensions emerging from the secondary target of H2 (and H4) select and transport the tertiary lower energy particles (mixed hadron, or almost pure electron beams within the range of 0.4 – 12 GeV) to the installed double phase in H2 and single phase ProtoDUNE experiments [2,3]. The beam layout and performance of the VLE beams are presented in detail in [1].

A CATIA drawing of the H2-VLE extension pointing to the 11×11 m$^2$ Double Phase ProtoDUNE experiment is shown in Fig.1.

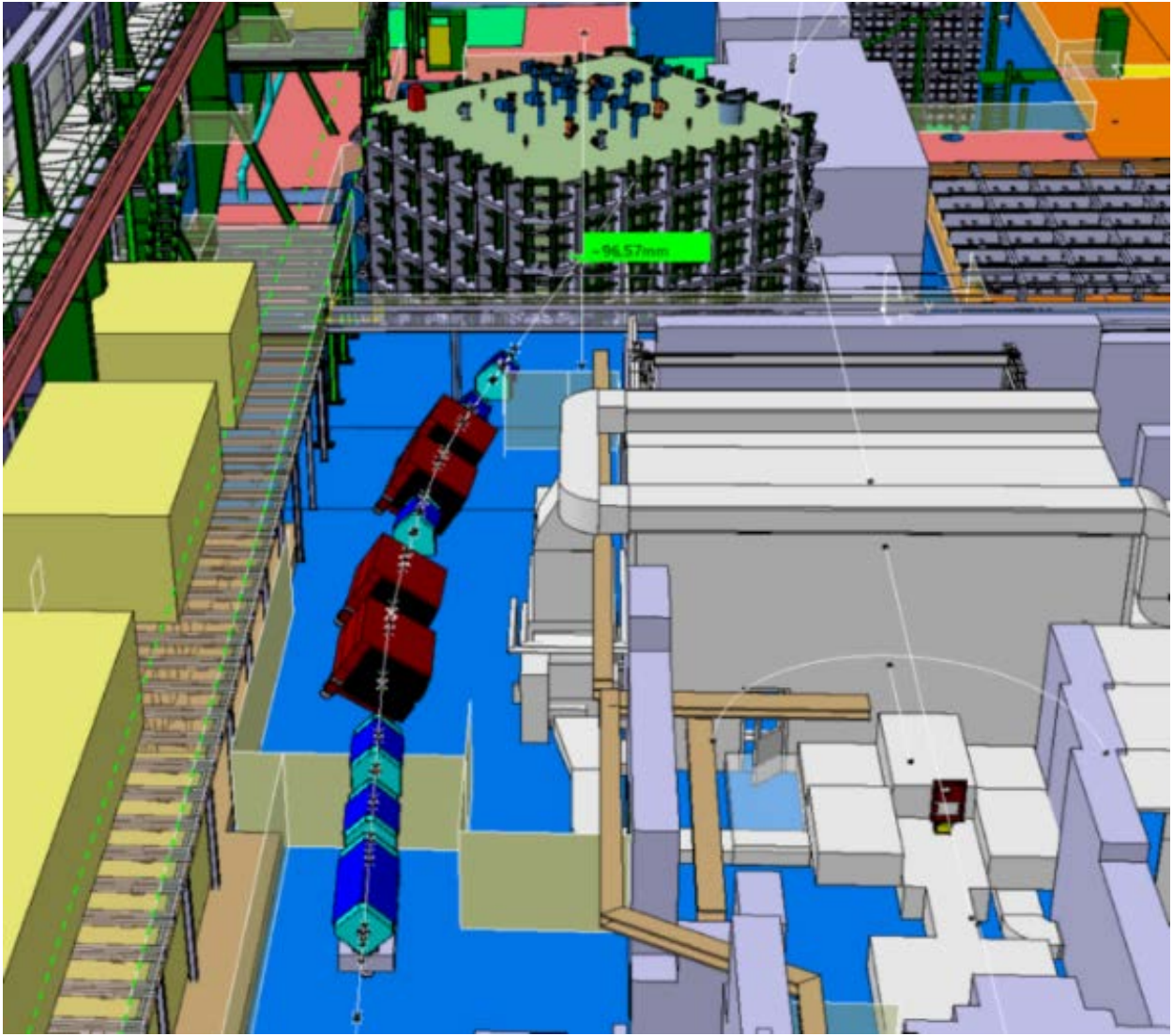

**Figure 1: CATIA drawing of the H2-VLE beam line extension. The total beam line length from the secondary target at the bottom of the figure to the experiment is 39.644 m. (Courtesy V. Clerc).**

In the case of H2-VLE beam line, downstream the secondary target the initial quadrupole acceptance triplet makes the first geometrical selection of the low energy particles, followed by a momentum selection and recombination station including the momentum collimator and the field-lens quadrupole. Finally, a set of focusing quadrupoles focuses the beam on the experiment, providing a spot size of approximately 10×6 cm$^2$ at the center of the detector. The conceptual layout of the beam line along with the instrumentation, is shown in Fig. 1. The maximum momentum spread accepted by the beam line is 5%. In order to maximize the acceptance and reduce the overall length of the beamline, the four bending magnets are tilted by 34.9 degrees with respect to the floor. The optics drawing of the beam line depicting the beam envelope is shown in Fig.2. Fig.3 shows the spot size of the beam at the focal point at the middle of the ProtoDUNE-DP detector.

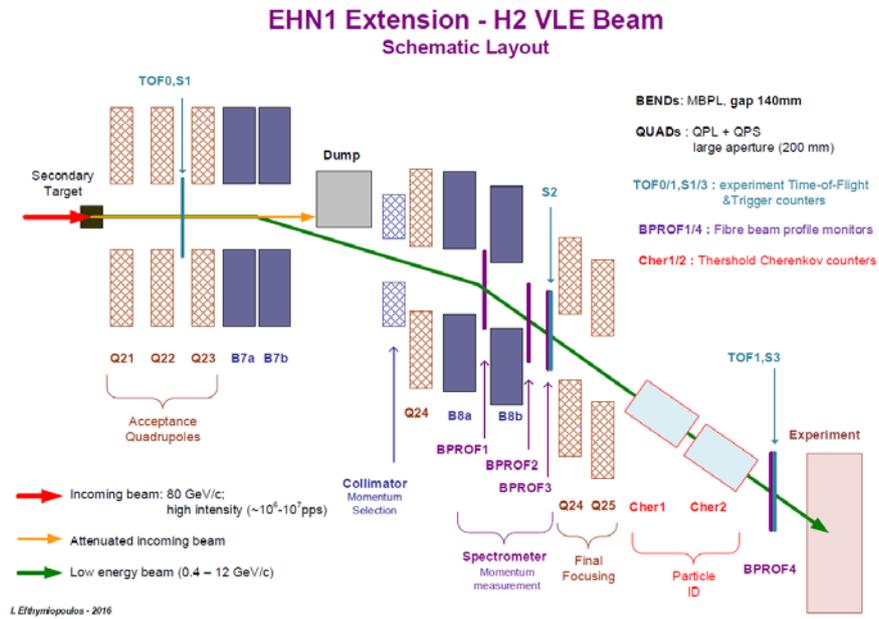

**Figure 1: Schematic layout of H2-VLE beam line. Between the elements, several instrumentation devices are depicted.**

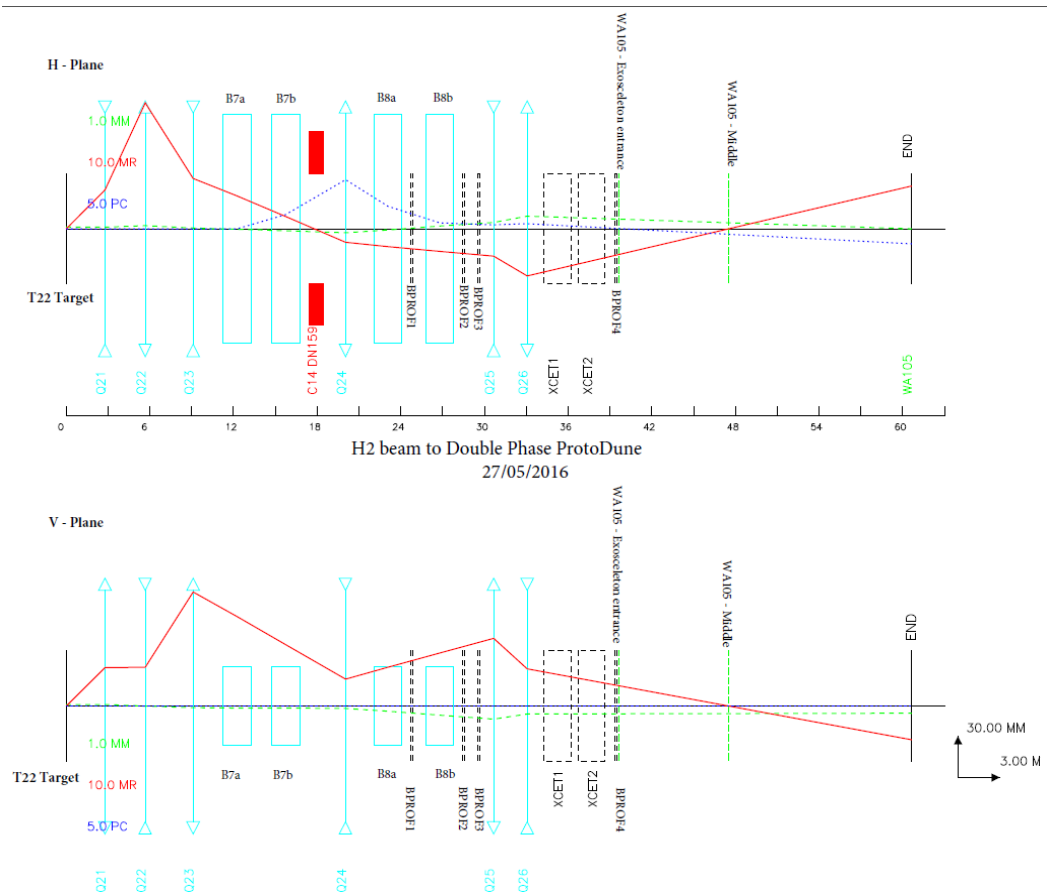

**Figure 2: Optics drawing of the H2-VLE beam line. The red line indicates the envelope of particles emitted at 10 mrad from the secondary target, the blue the contribution to the beam size from the momentum dispersion, and the green that of the target source of 1mm.**

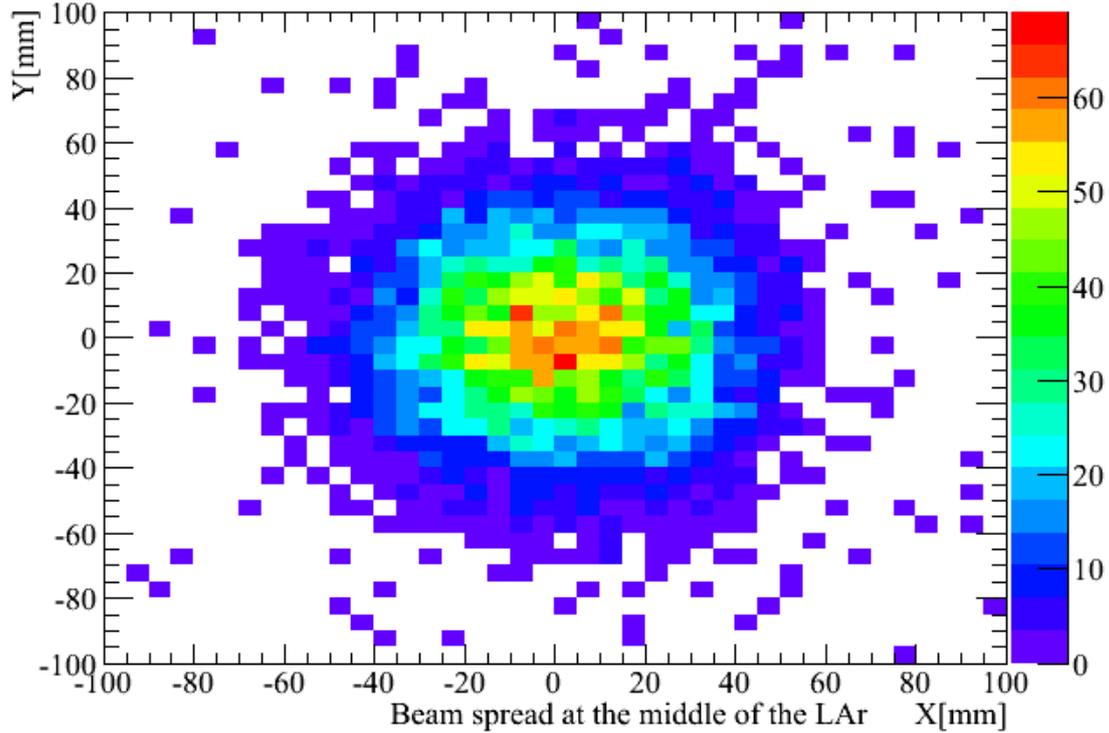

**Figure 3: Spot size of the beam at the middle of the protoDUNE-DP detector.**

## 2. Beam Composition and Rates

In order to investigate the beam particle composition and the background seen by the experiment and therefore optimize the particle identification methods and the beam instrumentation, a detailed model of the upstream secondary H2 beam line as well as the H2-VLE extension has been in a GEANT4 based software, G4BeamLine [4]. All the beam elements (magnets, collimators, vacuum elements and beam detectors) have been precisely modelled, and massive Monte Carlo simulation productions have been performed, in order to minimize the statistical uncertainties. The model takes into account all the physics processes included in GEANT4, eg: muon-to-gamma production synchrotron, radiation energy loss, multiple scattering, etc.

As first step, simulations were performed to validate the beam line design and explore different simulation options. In this case, the 'default' secondary target of a copper cylinder, 30 cm long and 3 cm radius was used. Further, to be as realistic as possible in terms of transported particles and intensities, the secondary particles were generated at the primary target and transported for 631 m through the magnetic elements of H2 beam line, interacting with the secondary target, and subsequently transported to the experiment through the VLE part. Two main physics lists from GEANT4 were used: the "FTFP_BERT" and "QGSP_BIC". Since the two lists depend on different parametrizations of the several hadronic cross-sections, the predicted production rate of particles can be quite different depending on the momentum range. A detailed study, comparing the simulation results with the test beam data should be done in order to finally assert the best physics list for this momentum range. The transport threshold of the particles was set to 10 MeV. A visualization of the model used in those studies can be seen in Fig. 5.

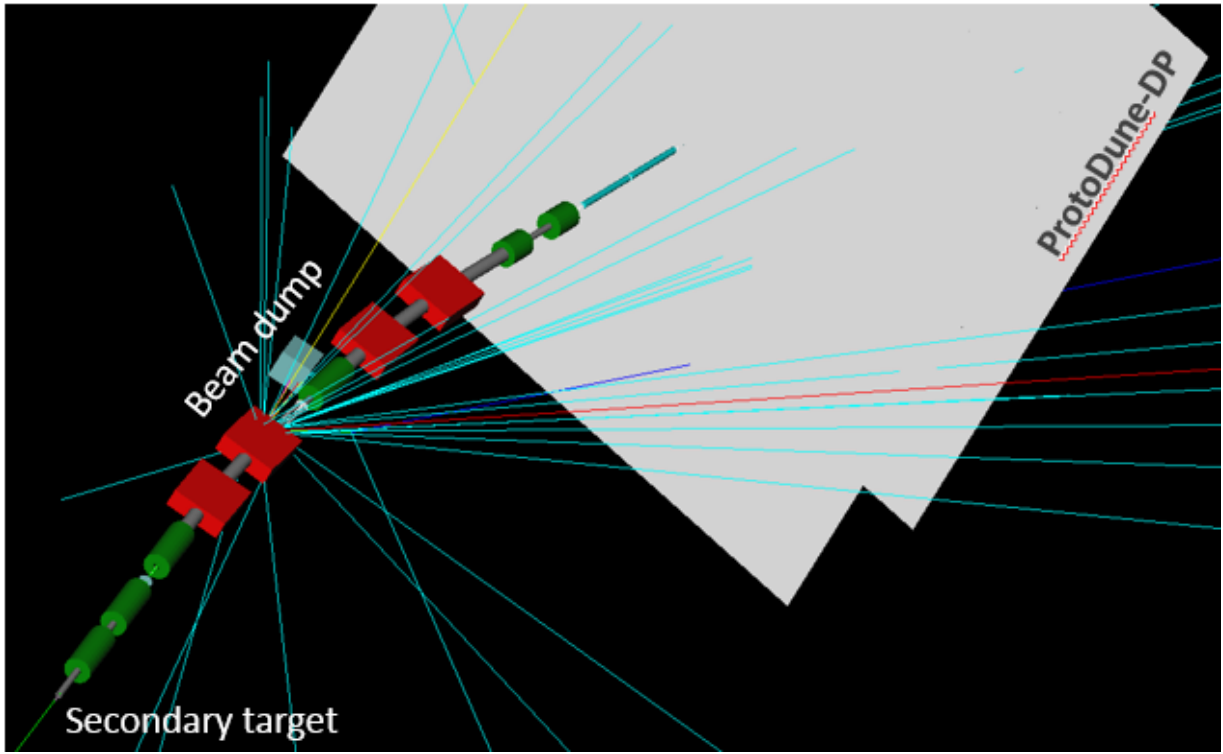

**Figure 5: A visualization of the H2-VLE beam line in G4BeamLine. The four tilted bends are depicted in red, the quadrupoles in green, and the ProtoDune-DP detector in gray. All the vacuum elements have been modeled, as well as the instrumentation. The blue tracks show the emerging low-energy particles from the interaction of the beam particles with at the magnets' apertures and yokes.**

Two different operational scenarii have been studied: The VLE beam line transports the *positive sign particles* or *the negative sign* particles respectively produced at the secondary target. For this analysis, we impose the condition that all particles at the entrance of the ProtoDUNE-DP detector, are within the aperture of the beam pipe, assumed rectangular, with horizontal and vertical dimensions of 200 mm. No additional condition has been imposed and therefore the beam composition may include low-energy particles generated from showers in the interaction of the beam particles with apertures along its length that would easily be rejected in the real experiment with the proper trigger setup.

The background to the experiment from particles outside the beam pipe is not discussed in this note, nor the muon background from the secondary beam (H2 or H4) or the neighboring beam lines or other sources. Also no shielding/dumps/walls or beam windows have been included in these simulations as well as no material from the instrumentation devices.

## 2.1 Hadron rate at the experiment

Fig.6 shows the full-acceptance useful hadronic rate of particles seen by the experiment as a function of the momentum for $10^6$ particles on the secondary target and the SPS spill length of 4.8 seconds. The experiment's DAQ is limited for a maximum rate of ~100 Hz to be provided by the beam, a condition that is satisfied for energies above 3-4 GeV, while for lower energies a higher secondary beam intensity must be used. The pion, kaon and proton content of the beam, normalized to the total charged particles, for both positive and negative momenta is shown in Figures 7, 8 and 9. Significant differences are observed between the two lists considered as for example in the pion production, where for the low-energies approx. 20% more pi$^-$ are produced compared to the positive case.

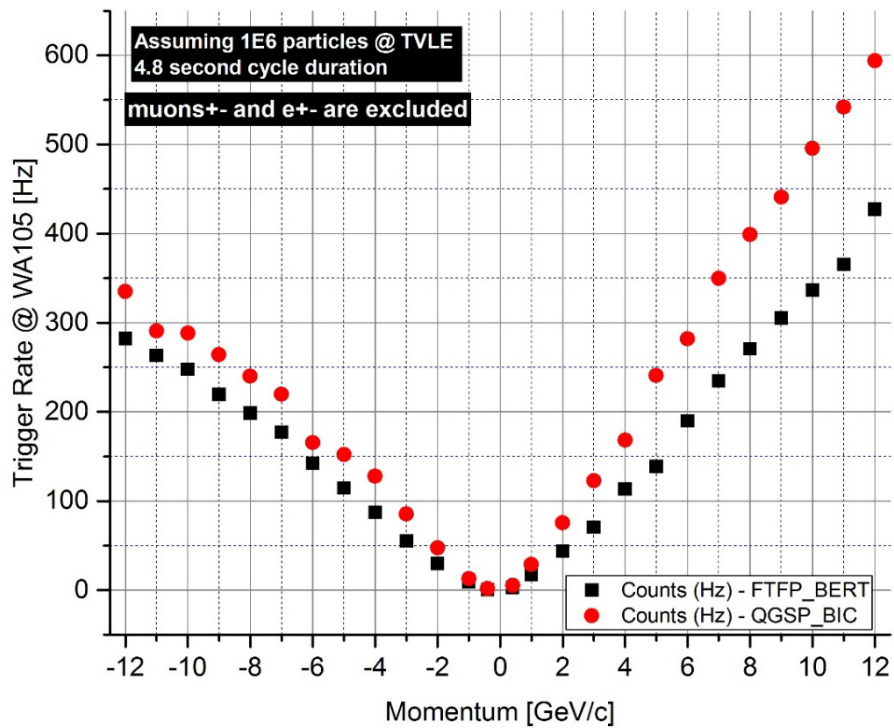

**Figure 6: Simulated hadron rate of all charged particles at ProtoDUNE-DP experiment, for 1E6 particles on the secondary target and a spill length of 4.8 s, using the "standard" target of 30 cm copper.**

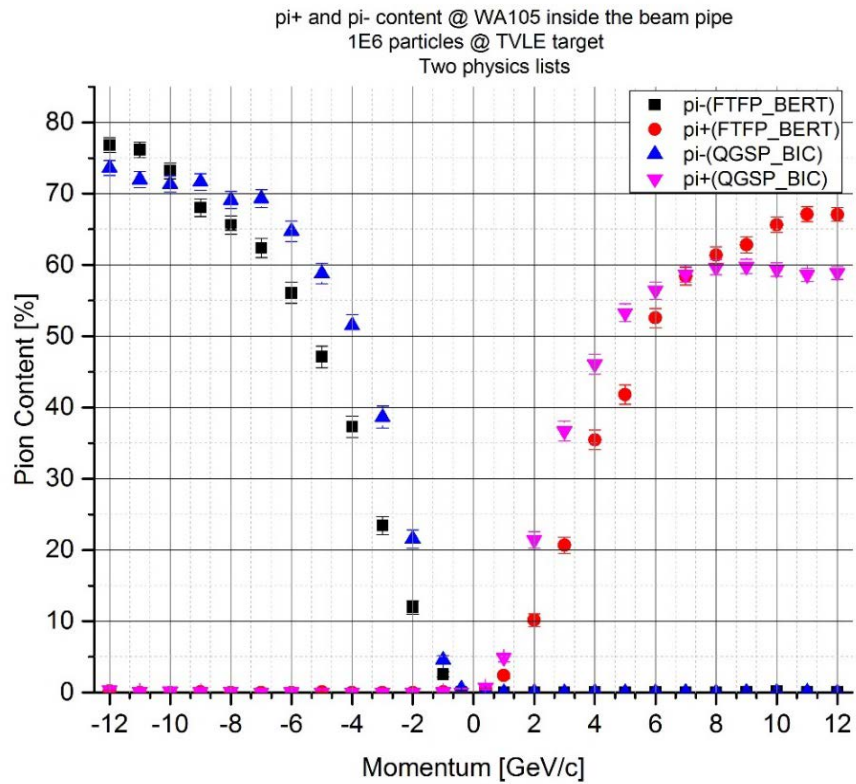

**Figure 7: Simulated pion content reaching the ProtoDUNE-DP experiment, for an initial intensity of $10^6$ particles on the secondary target. The errors are calculated using the binomial error function.**

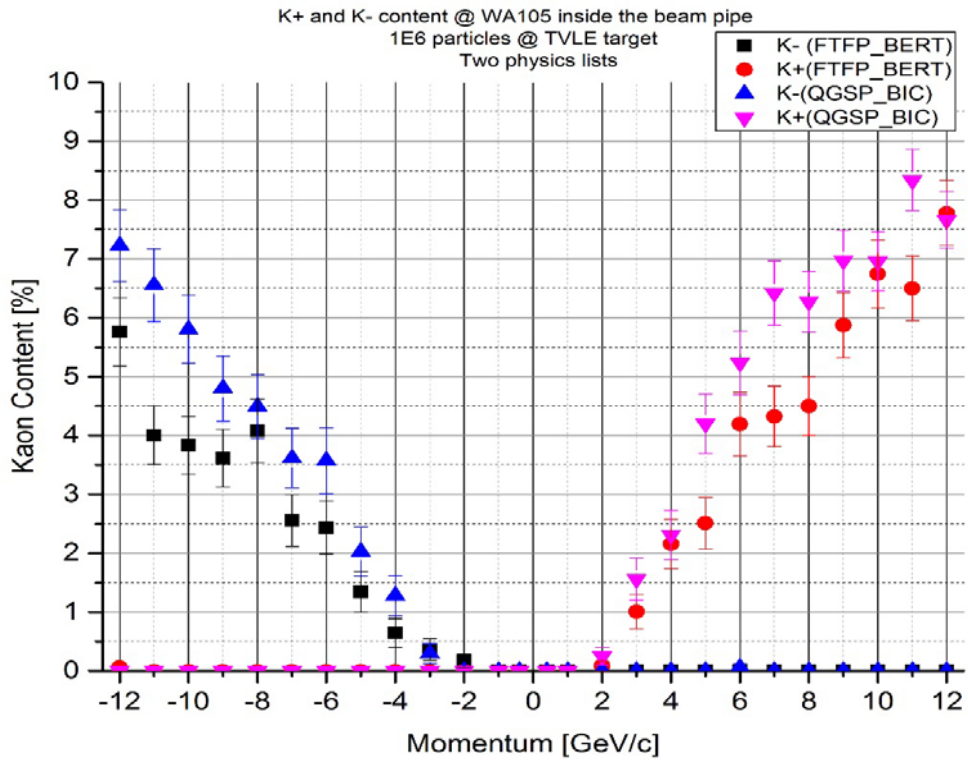

**Figure 8:** Simulated kaon content at ProtoDUNE-DP experiment, assuming an initial intensity of $10^6$ particles on the secondary target. The errors are calculated using the binomial error function.

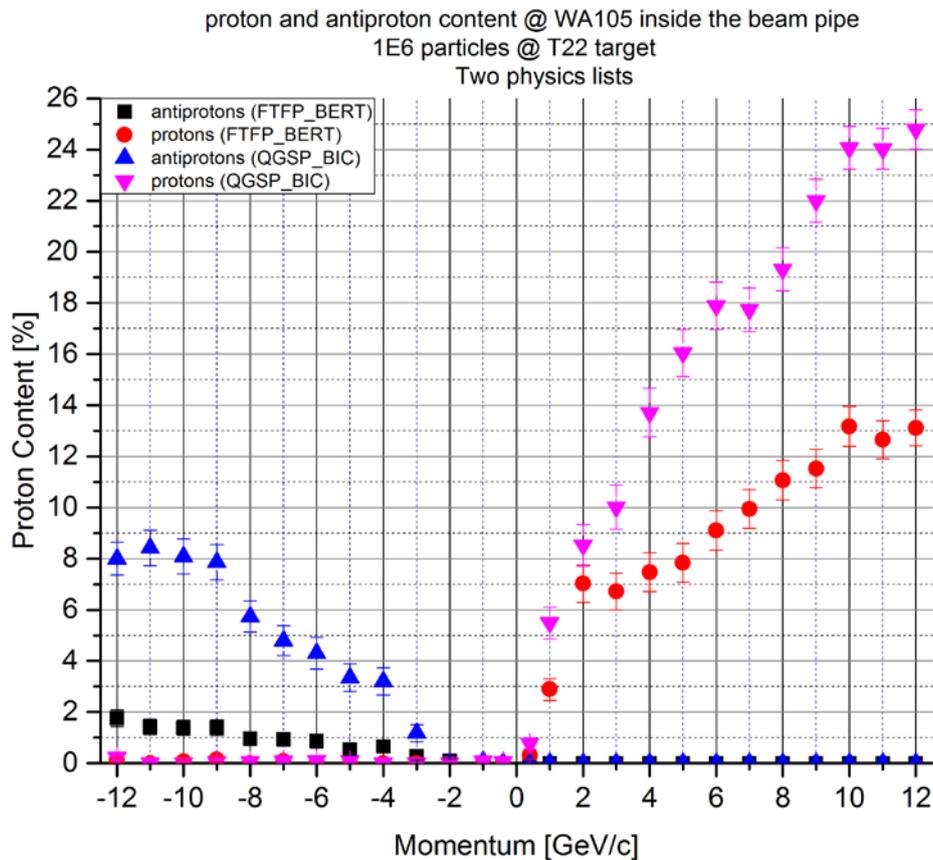

**Figure 9:** Simulated proton content at ProtoDUNE-DP experiment, assuming an initial intensity of $10^6$ particles on target. The errors are calculated using the binomial error function.

In the lower momenta, electrons dominate the spectrum. For the energies lower than 4 GeV, the decay length of pions and kaons becomes a significant factor in the population of those particles reaching the experiment. The percentage of surviving pions and kaons, as a function of the beam line length is shown in Figures 10 and 11. For the 0.4 GeV beam, approximately only 20 % of the initially produced pions will reach the entrance of the experiment, at 39. 644 m from the secondary target. For Kaons, below 4 GeV the expected rate to the experiment becomes negligible, and drops to zero below 2 GeV. Approximately 62% of the initially produced kaons with momentum 12 GeV will reach the experiment.

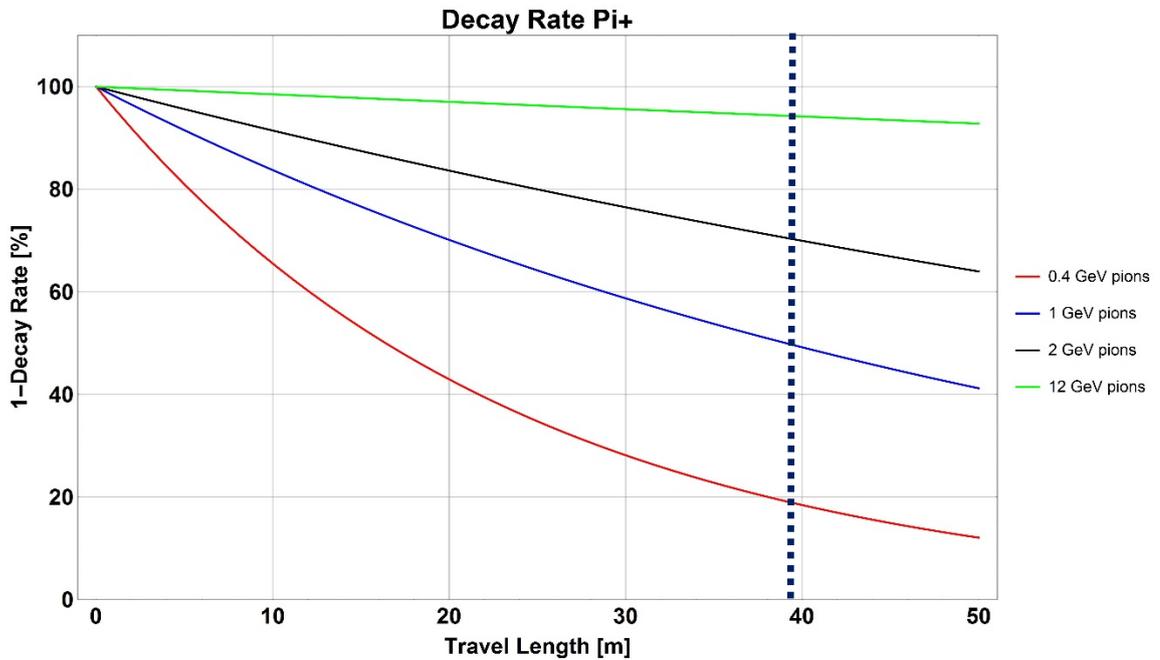

**Figure 10: Decay rate of pions as a function of the distance from the secondary target. The dashed vertical line indicates the entrance of the ProtoDUNE-DP experiment.**

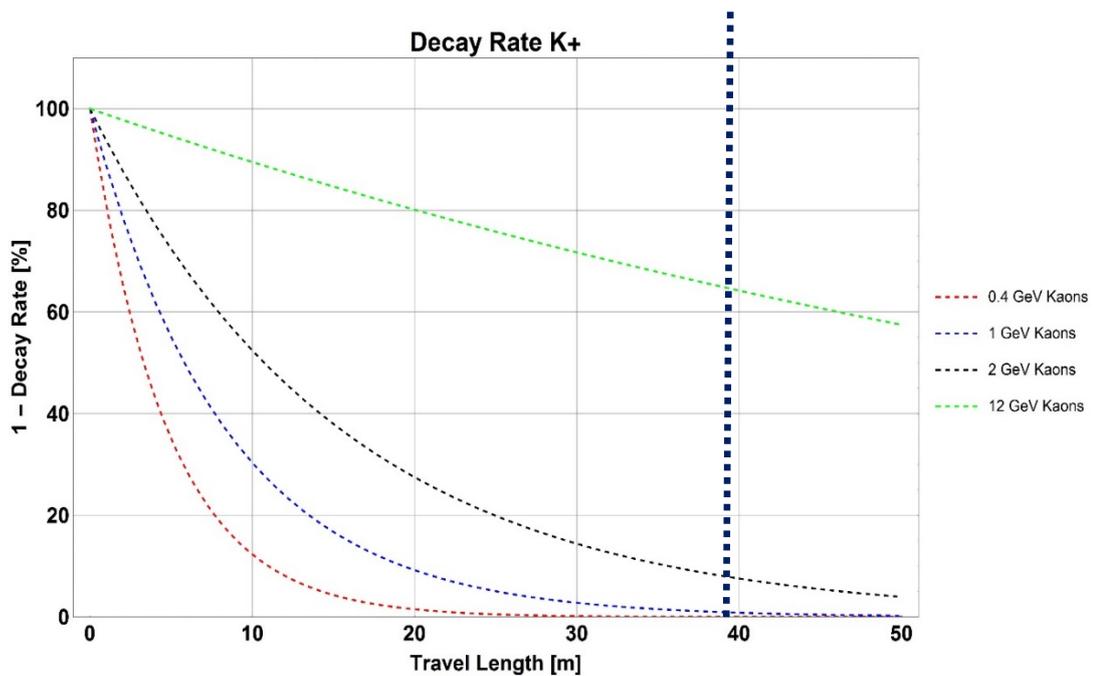

**Figure 11: Decay rate of kaons as a function of the distance from the secondary target. The dashed vertical line indicates the location of the ProtoDUNE-DP experiment.**

## 2.2 Hadron beam composition

The simulated beam content at the experiment, within the vacuum beam pipe aperture for both positive and negative particle beams is summarized in Tables 1 and 2, for the two different physics lists, and using the 'standard' Cu target. One million events have been simulated per energy point, resulting to an error in all cases of less than 3 %. Figures 12 and 13 show a graphical view of the same results.

| Momentum [GeV/c] | anti-p | e- | e+ | K- | K+ | mu- | mu+ | p | pi- | pi+ |
|---|---|---|---|---|---|---|---|---|---|---|
| -0,4 | 0.00% | 98.45% | 0.42% | 0.00% | 0.00% | 0.18% | 0.83% | 0.00% | 0.12% | 0.00% |
| -1 | 0.00% | 95.48% | 0.37% | 0.00% | 0.00% | 0.79% | 0.73% | 0.00% | 2.56% | 0.06% |
| -2 | 0.09% | 84.57% | 0.51% | 0.17% | 0.00% | 1.62% | 1.11% | 0.00% | 11.94% | 0.00% |
| -3 | 0.27% | 72.30% | 0.54% | 0.36% | 0.00% | 1.82% | 1.27% | 0.00% | 23.43% | 0.00% |
| -4 | 0.65% | 56.69% | 0.09% | 0.65% | 0.00% | 3.42% | 1.20% | 0.00% | 37.30% | 0.00% |
| -5 | 0.54% | 45.31% | 0.18% | 1.34% | 0.00% | 4.20% | 1.25% | 0.00% | 47.10% | 0.09% |
| -6 | 0.87% | 36.09% | 0.26% | 2.43% | 0.00% | 3.13% | 1.13% | 0.00% | 56.09% | 0.00% |
| -7 | 0.93% | 28.74% | 0.15% | 2.56% | 0.00% | 4.26% | 0.93% | 0.08% | 62.35% | 0.00% |
| -8 | 0.96% | 24.24% | 0.15% | 4.08% | 0.00% | 3.78% | 1.19% | 0.00% | 65.60% | 0.00% |
| -9 | 1.39% | 21.82% | 0.14% | 3.61% | 0.00% | 3.89% | 0.90% | 0.14% | 68.03% | 0.07% |
| -10 | 1.39% | 15.47% | 0.40% | 3.83% | 0.00% | 4.69% | 0.93% | 0.07% | 73.23% | 0.00% |
| -11 | 1.42% | 13.57% | 0.26% | 4.01% | 0.00% | 3.68% | 0.90% | 0.00% | 76.16% | 0.00% |
| -12 | 1.75% | 12.02% | 0.25% | 5.76% | 0.06% | 2.38% | 0.56% | 0.13% | 76.83% | 0.25% |
| 0,4 | 0.00% | 0.72% | 97.49% | 0.00% | 0.00% | 0.12% | 0.90% | 0.30% | 0.06% | 0.42% |
| 1 | 0.00% | 0.84% | 92.54% | 0.00% | 0.00% | 0.19% | 1.16% | 2.89% | 0.00% | 2.38% |
| 2 | 0.00% | 0.58% | 79.83% | 0.00% | 0.08% | 0.17% | 2.15% | 7.02% | 0.00% | 10.17% |
| 3 | 0.00% | 0.25% | 68.09% | 0.00% | 1.01% | 0.17% | 3.11% | 6.72% | 0.00% | 20.65% |
| 4 | 0.00% | 0.33% | 50.87% | 0.00% | 2.16% | 0.08% | 3.57% | 7.47% | 0.08% | 35.44% |
| 5 | 0.00% | 0.47% | 41.41% | 0.00% | 2.51% | 0.31% | 5.65% | 7.84% | 0.00% | 41.80% |
| 6 | 0.00% | 0.22% | 28.71% | 0.00% | 4.19% | 0.14% | 4.99% | 9.11% | 0.07% | 52.57% |
| 7 | 0.00% | 0.58% | 21.50% | 0.00% | 4.33% | 0.26% | 4.97% | 9.94% | 0.00% | 58.42% |
| 8 | 0.00% | 0.30% | 17.70% | 0.00% | 4.50% | 0.41% | 4.62% | 11.07% | 0.00% | 61.40% |
| 9 | 0.00% | 0.33% | 14.60% | 0.00% | 5.87% | 0.22% | 4.50% | 11.53% | 0.11% | 62.84% |
| 10 | 0.00% | 0.42% | 10.20% | 0.00% | 6.74% | 0.05% | 3.56% | 13.17% | 0.21% | 65.64% |
| 11 | 0.00% | 0.34% | 9.06% | 0.00% | 6.50% | 0.05% | 4.19% | 12.65% | 0.10% | 67.11% |
| 12 | 0.00% | 0.39% | 8.08% | 0.00% | 7.78% | 0.09% | 3.35% | 13.11% | 0.13% | 67.08% |

**Table 1: VLE beam particle content at the experiment, within the beam pipe and for 1E6 particles impinging on the secondary target. The physics list used is FTFP_BERT.**

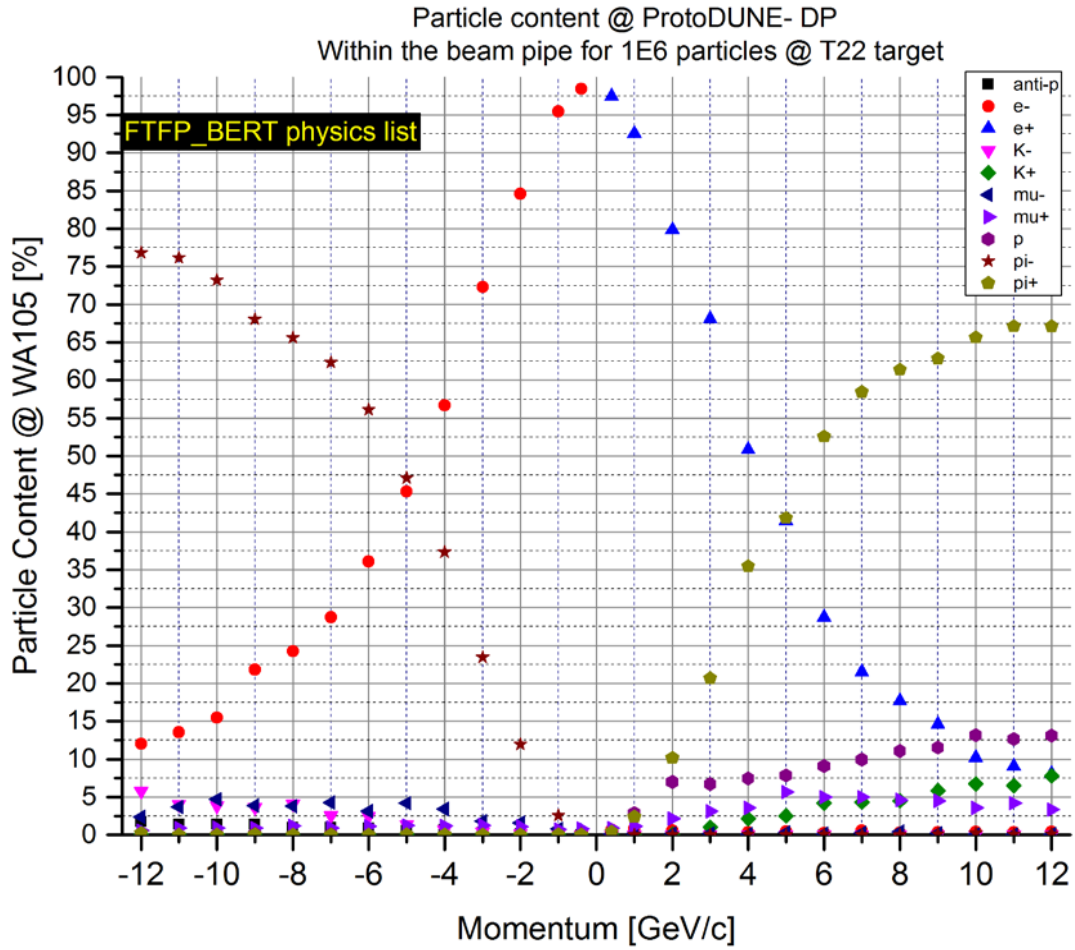

**Figure 12:** Particle content at the ProtoDune-DP experiment, within the beam pipe aperture, for $10^6$ particles impinging on the secondary target. The physics list used is FTFP_BERT.

| Momentum [GeV/c] | anti-p | e- | e+ | K- | K+ | mu- | mu+ | p | pi- | pi+ |
|---|---|---|---|---|---|---|---|---|---|---|
| -0,4 | 0.00% | 97.62% | 0.60% | 0.00% | 0.00% | 0.20% | 0.99% | 0.07% | 0.53% | 0.00% |
| -1 | 0.08% | 92.20% | 0.47% | 0.00% | 0.00% | 1.17% | 1.33% | 0.08% | 4.60% | 0.08% |
| -2 | 0.00% | 73.88% | 0.19% | 0.00% | 0.00% | 2.85% | 1.52% | 0.00% | 21.56% | 0.00% |
| -3 | 1.17% | 53.91% | 0.39% | 0.29% | 0.00% | 3.91% | 1.66% | 0.00% | 38.65% | 0.00% |
| -4 | 3.20% | 36.71% | 0.09% | 1.28% | 0.00% | 5.75% | 1.46% | 0.00% | 51.51% | 0.00% |
| -5 | 3.35% | 28.02% | 0.26% | 2.03% | 0.00% | 6.61% | 0.88% | 0.09% | 58.77% | 0.00% |
| -6 | 4.30% | 22.16% | 0.18% | 3.57% | 0.00% | 3.66% | 1.19% | 0.09% | 64.74% | 0.09% |
| -7 | 4.79% | 16.67% | 0.37% | 3.61% | 0.00% | 3.98% | 1.18% | 0.07% | 69.32% | 0.00% |
| -8 | 5.74% | 15.34% | 0.14% | 4.49% | 0.00% | 3.80% | 1.17% | 0.07% | 69.11% | 0.14% |
| -9 | 7.87% | 11.47% | 0.07% | 4.80% | 0.00% | 2.87% | 1.13% | 0.07% | 71.67% | 0.07% |
| -10 | 8.09% | 9.51% | 0.25% | 5.81% | 0.00% | 3.83% | 0.99% | 0.00% | 71.34% | 0.19% |
| -11 | 8.43% | 8.49% | 0.12% | 6.55% | 0.00% | 3.12% | 1.19% | 0.00% | 71.97% | 0.12% |
| -12 | 8.00% | 6.11% | 0.11% | 7.22% | 0.00% | 3.78% | 0.61% | 0.22% | 73.61% | 0.33% |
| 0,4 | 0.00% | 0.52% | 96.65% | 0.00% | 0.00% | 0.06% | 1.16% | 0.77% | 0.13% | 0.71% |
| 1 | 0.00% | 0.30% | 87.06% | 0.00% | 0.00% | 0.38% | 1.88% | 5.49% | 0.00% | 4.89% |
| 2 | 0.00% | 0.58% | 64.33% | 0.00% | 0.25% | 0.08% | 4.76% | 8.52% | 0.08% | 21.39% |
| 3 | 0.00% | 0.16% | 46.47% | 0.00% | 1.56% | 0.25% | 4.84% | 10.02% | 0.00% | 36.70% |
| 4 | 0.00% | 0.31% | 31.51% | 0.00% | 2.31% | 0.08% | 5.93% | 13.71% | 0.08% | 46.07% |
| 5 | 0.00% | 0.38% | 20.56% | 0.00% | 4.20% | 0.06% | 5.47% | 16.04% | 0.00% | 53.28% |
| 6 | 0.00% | 0.18% | 14.65% | 0.06% | 5.24% | 0.18% | 5.35% | 17.88% | 0.06% | 56.41% |

| | | | | | | | | | |
|---|---|---|---|---|---|---|---|---|---|
| 7  | 0.00% | 0.40% | 11.66% | 0.00% | 6.42% | 0.15% | 4.89% | 17.74% | 0.05% | 58.70% |
| 8  | 0.00% | 0.18% | 10.24% | 0.00% | 6.28% | 0.00% | 4.36% | 19.31% | 0.00% | 59.64% |
| 9  | 0.00% | 0.21% | 7.39%  | 0.00% | 6.97% | 0.04% | 3.53% | 22.00% | 0.08% | 59.78% |
| 10 | 0.00% | 0.23% | 5.63%  | 0.00% | 6.96% | 0.08% | 3.61% | 24.07% | 0.08% | 59.35% |
| 11 | 0.00% | 0.35% | 5.08%  | 0.00% | 8.34% | 0.07% | 3.40% | 24.03% | 0.11% | 58.63% |
| 12 | 0.00% | 0.16% | 4.65%  | 0.00% | 7.66% | 0.06% | 3.75% | 24.78% | 0.03% | 58.90% |

**Table 2: Particle content at the experiment, within the beam pipe aperture, for $10^6$ particles impinging on the secondary target. The physics list used is QGSP_BIC.**

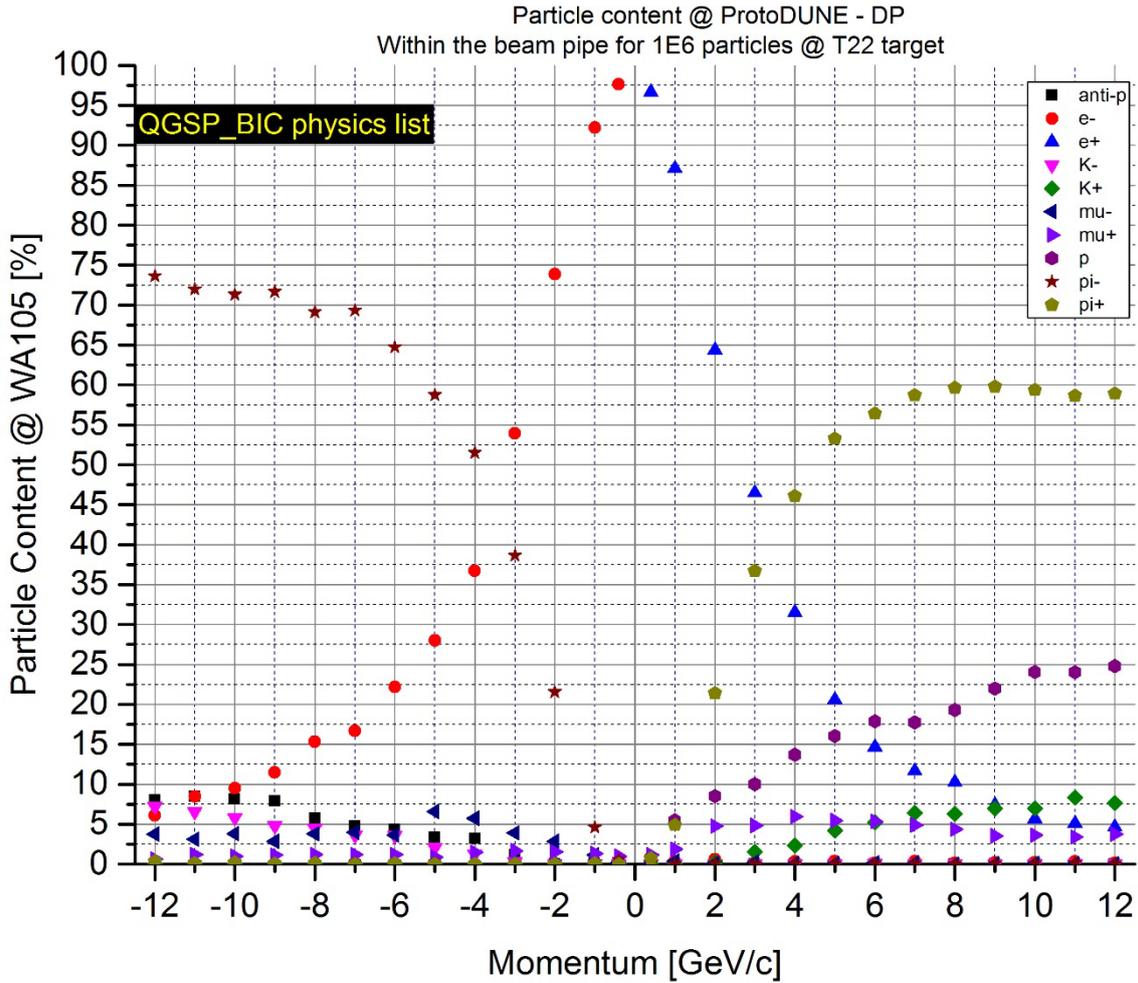

**Figure 13: Particle content at the experiment, within the beam pipe, for $10^6$ particles impinging on the secondary target. The physics list used is QGSP_BIC.**

From the above tables and plots, is evident that the difference in the predictions of the physics lists used, becomes more pronounced in the lower energy part of the spectrum, and more specifically in the anti-proton production. The ratio of the particle counts in the experiment between the two lists is shown in Figure 14. The ratio of all the particles species remains within a factor of 2.5, with the exception of the big overestimation by the QGSP_BIC list for the antiproton production in the higher energies by a maximum factor of 6.5. Several references (see for example [5]) recommend the FTFP_BERT list for simulating the hadron production in the 1-12 GeV momentum range, and therefore we have chosen this one as baseline for the rest of the studies.

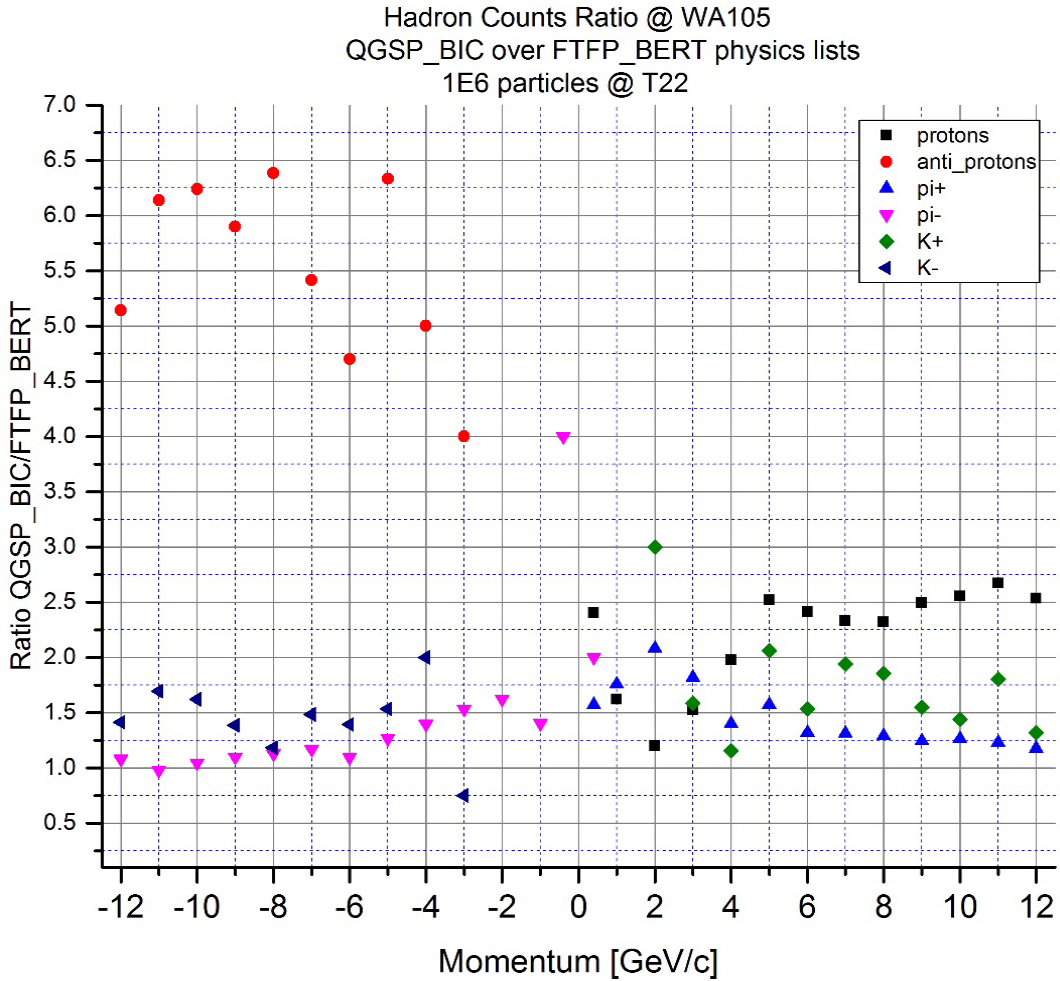

**Figure 14: Ratio of hadron counts at ProtoDune-DP of the two physics lists used in the study.**

## 2.3 Target optimization for hadron beams

In order to enhance the pion production in the momentum range lower than 4 GeV/c, different target materials and dimensions have been studied. As shown in Table 1 and Figure 12, using as an example the 2 GeV/c case, the particle content reaching the experiment is ~ 80% electrons, ~10% pions and ~7% protons, with the standard Cu target. Changing the target material to tungsten and maintaining the same length of 30 cm and radius of 3 cm, the radiation length is increased by a factor ~5, and therefore the electron production is suppressed due to reabsorption. The pion production on W is slightly less than Cu, but the overall pion content of the beam is increased by a factor ~3, as shown in Figure 15. Figure 16 shows the dependence of the particle production of hadrons (pions and protons) on the target parameters: radius and length for the low-energy points of 1 and 2 GeV.

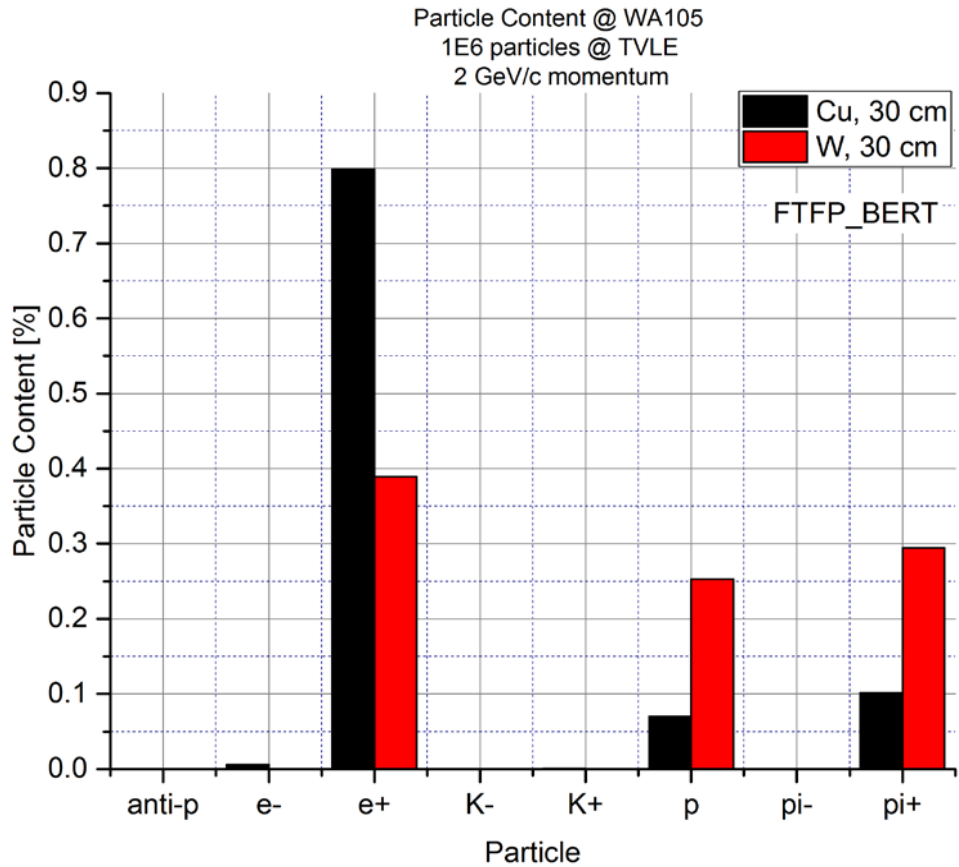

**Figure 15: Particle content with W and Cu Target, 30 cm length, for a 2 GeV/c beam.**

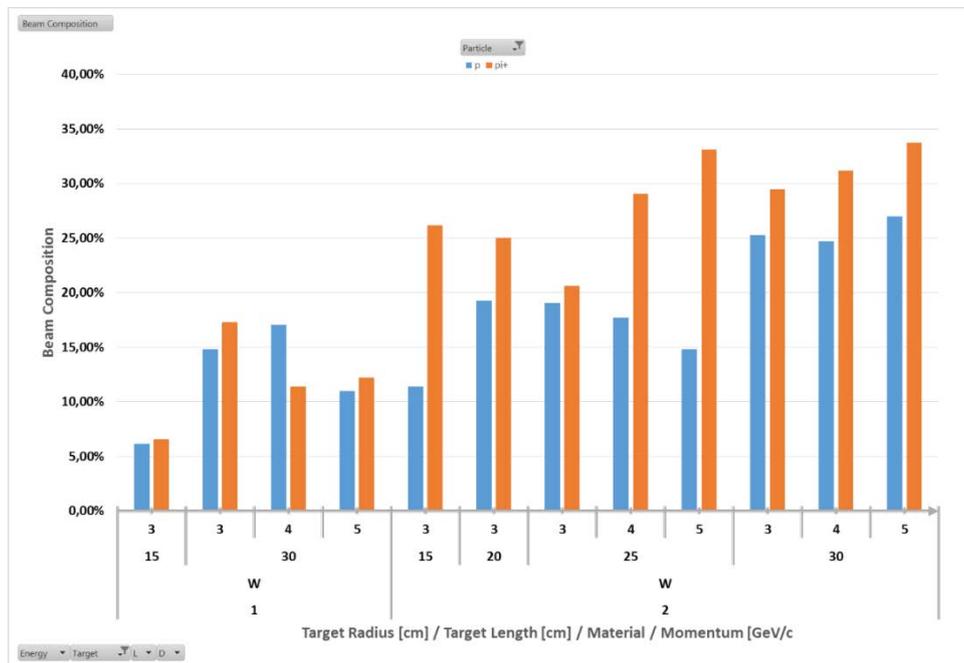

**Figure 16: Particle production from a W target of several lengths (15 to 30 cm) and radii (from 3 to 5 cm) for the two low-energy momenta 1 and 2 GeV.**

A W target of 30 cm length and 3 cm radius gives satisfactory yield for the 1 and 2 GeV/c momenta, and therefore we propose its use in the lower part of the momentum spectrum. For the higher energies the "standard" Cu target, of 30 cm length and 3 cm radius yields satisfactory results and therefore maintained.

## 2.4 Electron or muon beams

As discussed above, with the VLE beam set for hadron production and transport, we end up at the experiment with mixed beam containing hadrons, electrons and muons at variable proportions depending on the energy. The beam line will be equipped with special instrumentation as explained in Section 4 aimed to tag each particle species, in particular the pions and kaons that are more challenging for the low-energy range. Alternatively, low-energy electron or muon beams can be easily configured that represent much less of a challenge to achieve at the wanted trigger rate of the experiment of 100 Hz.

For the electron (or positron) beams the H2 (or H4) beam line will be tuned to transport a medium-energy secondary electron beam of 50-60 GeV of moderate intensity of approx. $10^4$ particles per pulse. For the low-energy production a Pb target of few radiation lengths will be used, tuned to the required energy range. The tertiary VLE beam will be basically pure, with the only background coming from low-energy electrons/photons from interaction with the beam aperture and material that can be cleared with the experimental trigger setup and the wanted electrons tagged with installed instrumentation.

Low-energy muon beams can be produced in the hadron mode by closing the momentum selection collimator. In this case, the last two dipoles in the line will act as momentum spectrometer to transport the selected beam momenta to the experiment.

# 3. Triggered Beam Composition and Data Taking Time Estimate

Based on the conclusions of Section 2, we finalized the beam line model inside GEANT4 and we run a high statistics simulation of 6M events per energy point, including all known materials along the beam: scintillators, Cherenkov gases, vacuum windows, stainless steel cryostat membrane. In addition we request that particles satisfy the minimal trigger conditions of the experiment, namely that they cross all the beam detectors. Table 3 shows the expected beam composition. Negative particle contribution is negligible.

| Momentum [GeV/c] | e+ | K+ | mu+ | p | pi+ |
|---|---|---|---|---|---|
| 0.4 | 97.61% | 0.00% | 0.00% | 0.48% | 1.91% |
| 1 | 74.20% | 0.00% | 0.00% | 14.94% | 10.86% |
| 2 | 45.83% | 0.67% | 0.96% | 20.04% | 32.50% |
| 3 | 25.16% | 1.56% | 1.56% | 17.56% | 54,17% |
| 3 | 68.29% | 0.64% | 0.42% | 7.72% | 22.94% |
| 4 | 53.72% | 1.46% | 0.65% | 7.56% | 36.61% |
| 5 | 42.38% | 2.47% | 0.83% | 9.18% | 45.14% |
| 6 | 31.42% | 3.83% | 0.73% | 10.10% | 53.92% |
| 7 | 24.70% | 4.08% | 0.85% | 9.92% | 60.46% |
| 8 | 19.36% | 5.11% | 0.97% | 11.33% | 63.24% |
| 9 | 15.12% | 5.67% | 0.82% | 11.10% | 67.29% |
| 10 | 12.36% | 6.02% | 0.71% | 12.25% | 68.66% |
| 11 | 10.46% | 6.95% | 0.82% | 13.57% | 68.20% |
| 12 | 8.90% | 6.89% | 0.66% | 14.26% | 69.30% |

**Table 3: Beam composition for the hadron beams, including all the material in the beam line. Below and at 3 GeV the W-target was used.**

Considering the requested number of particles for physics by the ProtoDUNE-DP experiment [6], we compute the number of days required to achieve this statistics with conservative estimates assuming: 50% useful events, an SPS supercycle of 50 s with two beam pulses to the North Area each of 4.8 seconds, and a trigger rate of 100 Hz for the experiment that based on the discussion in Section 2 can be reached for all energies. The results are shown in Table 4. The total number of run-days needed to achieve the experiment's goal is 84.

| Momentum [GeV/c] | Pion composition | Number of days | Number of Events |
|---|---|---|---|
| 0.4 | 1.91% | 32 | 500000 |
| 1 | 10.86% | 6 | 500000 |
| 2 | 32.50% | 4 | 1000000 |
| 3 | 22.94% | 11 | 2000000 |
| 4 | 36.61% | 7 | 2000000 |
| 5 | 45.14% | 5 | 1500000 |
| 6 | 53.92% | 4 | 1500000 |
| 7 | 60.46% | 3 | 1500000 |
| 8 | 63.24% | 3 | 1500000 |
| 9 | 67.29% | 3 | 1500000 |
| 10 | 68.66% | 3 | 1500000 |
| 11 | 68.20% | 3 | 1500000 |
| 12 | 69.30% | 3 | 1500000 |

**Table 4: Number of days needed in order to gather the required statistics for the ProtoDUNE-DP experiment, for each momentum bin according to the operational scenario as described in the text.**

The trigger rate, assuming a wide-open collimator and $10^6$ particles on the secondary target is shown in Tables 5 and 6 below.

| Momentum | e+ | K+ | mu+ | p | pi+ | Trigger rate [Hz] |
|---|---|---|---|---|---|---|
| 0.4 | 7 | 0 | 0 | 0 | 0 | 7 |
| 1 | 21 | 0 | 0 | 4 | 3 | 28 |
| 2 | 17 | 0 | 0 | 7 | 12 | 36 |
| 3 | 14 | 1 | 1 | 10 | 30 | 56 |

**Table 5: Trigger rate for a W target**

| Momentum | e+ | K+ | mu+ | p | pi+ | Trigger rate [Hz] |
|---|---|---|---|---|---|---|
| 3 | 145 | 1 | 1 | 16 | 49 | 213 |
| 4 | 117 | 3 | 1 | 16 | 80 | 218 |
| 5 | 94 | 5 | 2 | 20 | 100 | 222 |
| 6 | 77 | 9 | 2 | 25 | 133 | 247 |
| 7 | 69 | 11 | 2 | 28 | 169 | 279 |
| 8 | 59 | 16 | 3 | 35 | 193 | 305 |
| 9 | 51 | 19 | 3 | 37 | 227 | 337 |
| 10 | 46 | 22 | 3 | 45 | 254 | 370 |
| 11 | 41 | 27 | 3 | 53 | 268 | 393 |
| 12 | 38 | 29 | 3 | 60 | 292 | 422 |

**Table 6: Trigger rate for a Cu target**

## 4. Beam Instrumentation

The VLE beam line will be equipped with several types of beam instrumentation devices with a goal to: a) provide the beam profiles at various locations to be used for beam tuning, b) provide a trigger to the experiment, c) measure, particle by particle the beam momentum with a precision lower than 5% over the total operational momentum spectrum (0.4 GeV– 12 GeV) and finally d) provide particle identification, tagging pions, kaons, protons and electrons when possible. The various signals and information from the beam devices will be integrated with the experiment's DAQ system for synchronous readout.

We propose the following detectors along the beam line:

a) **4 "beam profile monitors"**, each one consisting of one or two layers of thin scintillating fibers of 1 mm width, measuring the horizontal and vertical coordinates (only in the front of the experiment). They will be used for the beam profile measurements and the event-by-event momentum determination. The four profile monitors are located at distances 24.882, 25.246, 29. 668 and 39.492 m from the secondary target. They are marked as "BPROFx", with x=1,2,3,4 in Figure 3.

b) **Two gas threshold Cherenkov counters**, each one with an active length of approx. two meters, located at distances 36.738 and 38.638 m from the secondary target, shown in Figure 3 as "XCET1" and "XCET2". The counters will be filled with Freon 12 and $CO_2$ gases, and their pressure is adjustable from ~0.5 bar up to 15 bar (for the $CO_2$ only) depending on the beam momentum.

c) **One extra fiber layer, which, in coincidence with one layer of BPROF4 will act as Time of Flight counter.** It is located at a distance of 7.304 m from the target (between Q22 and Q23), thus providing a flight distance of approx. 32 m, and

d) **Three thin scintillator tiles**, used to build the experiment's trigger. Using the aforementioned instrumentation, the momentum measurement method and the particle identification procedure is described in detail in the following sections.

All fiber and scintillator detectors will be installed inside the beam vacuum, thus minimizing the material in the beam. The two Cherenkov counters evidently must be separated and isolated from the beam vacuum with thin Mylar windows.

### 4.1 Momentum measurement

The theoretical momentum acceptance of H2-VLE (and H4-VLE) beam line is ~5%. The beam is equipped with a momentum selection station (including the 'momentum collimator') able to reduce the momentum byte Δp/p rms of the beam according to the empirical formula valid for this beam line:

$$\frac{\Delta p}{p} = \frac{\sqrt{C_x^2 + 51.1^2}}{22.6}\%$$

Where $C_x$ is the full opening of the collimator in mm. Figure 17 shows the Δp/p variation as a function of the collimator slit.

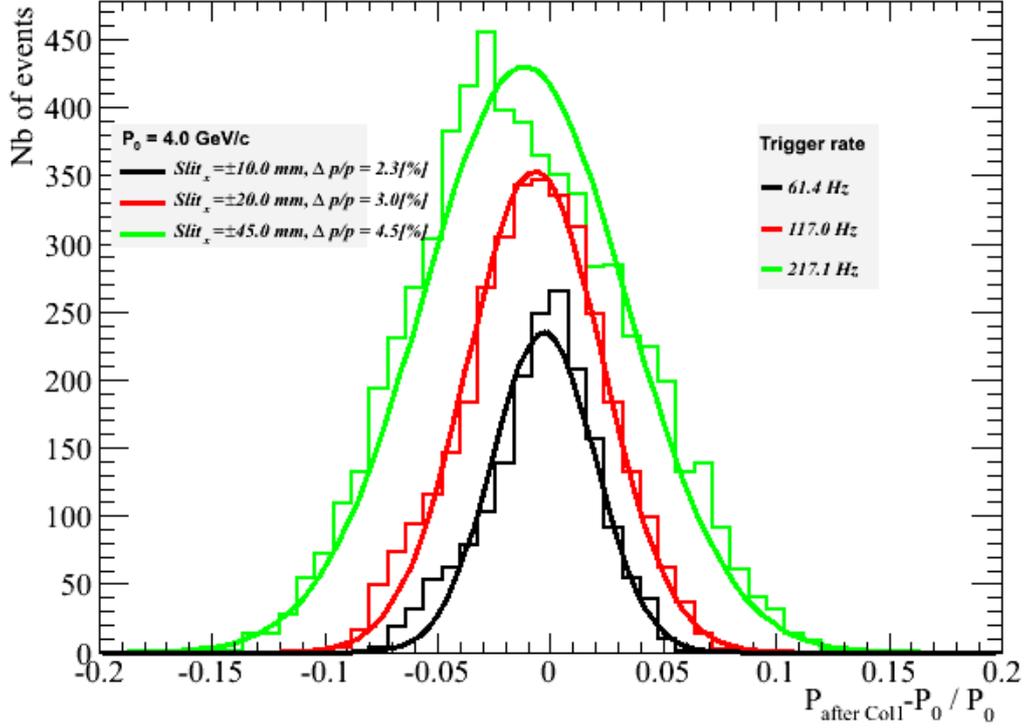

Figure 17: The beam momentum byte (Δp/p) for few settings of the collimator slit (GEANT4 simulation)

To achieve a better momentum resolution without reducing the count rate, the three beam profile monitors described in the previous section, placed around the fourth bending magnet (B8b) of the line are used as a spectrometer. The technique has been used in the past with success [6]. The layout of the spectrometer is shown in Figure 18. The distances of the three profile monitors from the center of the last magnet are 2.004 m, 1.718 m and 2.728 m respectively for BPROF1, BPROF2 and BPROF3, from which the distances $L_1$, $L_2$, and $L_3$ relevant for the momentum reconstruction can be deduced. In a simplified approach, the deflection angle θ of a beam particle traversing the magnetic field of the dipole can be determined using the measured position in the three monitors P1, P2, P3 and the point Po that is the projection of the outgoing trace at the origin as shown in Figure 18. Using vector notation, we have:

$$\cos\vartheta = \frac{\overrightarrow{P_1P_o} \cdot \overrightarrow{P_3P_2}}{\|P_1P_o\|\|P_3P_2\|}$$

Which after some algebra gives:

$$\cos\theta = \frac{(a-x_1)(x_3-x_2)\cos\theta_0 - (L_3-L_2)[L_1+(a-x_1)\tan\theta_0]}{\sqrt{L_1^2+(a-x_1)^2}\sqrt{(L_3-L_2)^2+[(L_3-L_2)\tan\theta_0-(x_3-x_2)\cos\theta_0]^2})}$$

$$a = \frac{x_2L_3 - x_3L_2}{L_3 - L_2}$$

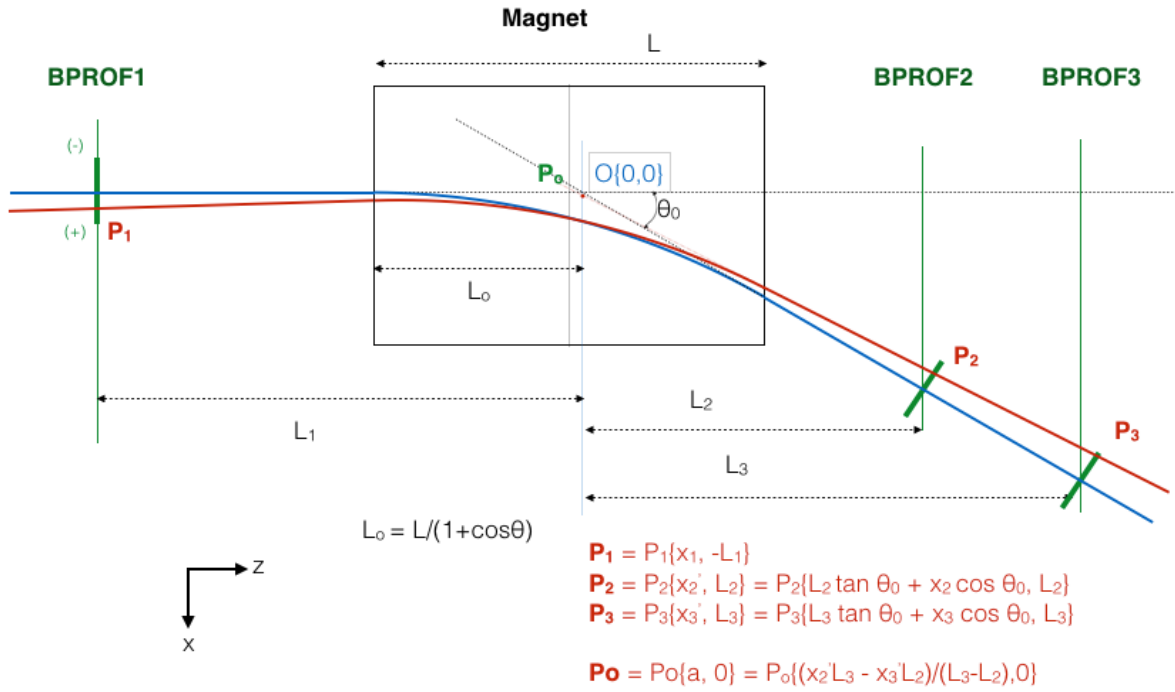

**Figure 18: Layout of the H2-VLE (and similarly H4-VLE) momentum spectrometer around the last dipole.**

To validate the performance of the spectrometer, we used the high statistics simulation, which includes all the material in the line, the gas in the Cherenkov detectors at the right pressures per momentum, as well as the expected special resolution of the profile monitors. For each particle, we compute its momentum from the above equation, and therefore the measured $\Delta p/p$ of the line. Assuming no material in the beam line for a central momentum of 12 GeV/c and position resolutions of 0.2 mm, 0.5 mm and 0.8 mm we obtain a $\Delta p/p$ of 1.1%, 2.5% and 3.9% accordingly, as shown on Figure 19. When the material along the beam is included, the reconstructed momentum resolution $\Delta p/p$ deteriorates, because of the multiple scattering, with the effect becoming more significant in lower energies, as shown on Figures 20, 21 and 22. For the 2 GeV beam, the reconstructed momentum resolution with all material included and with 0.2mm position resolution in the detectors increases to 1.5% and gets to about 6% for the lowest energy point of 0.4 GeV.

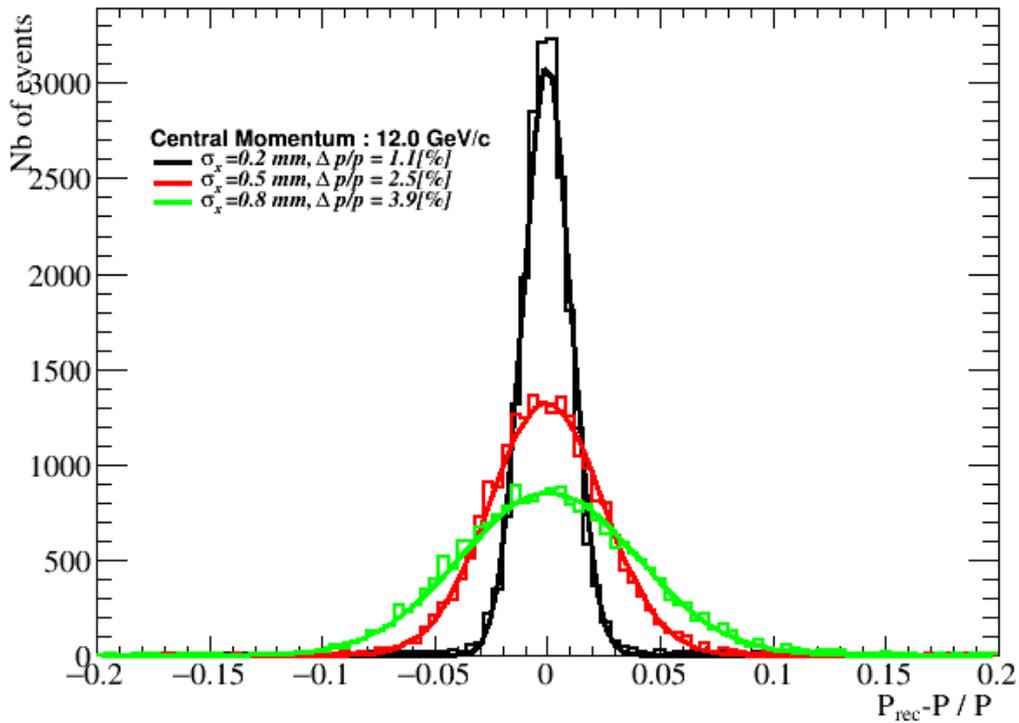

**Figure 19: Reconstructed momentum byte for a beam of 12 GeV, ignoring the material on the beam line.**

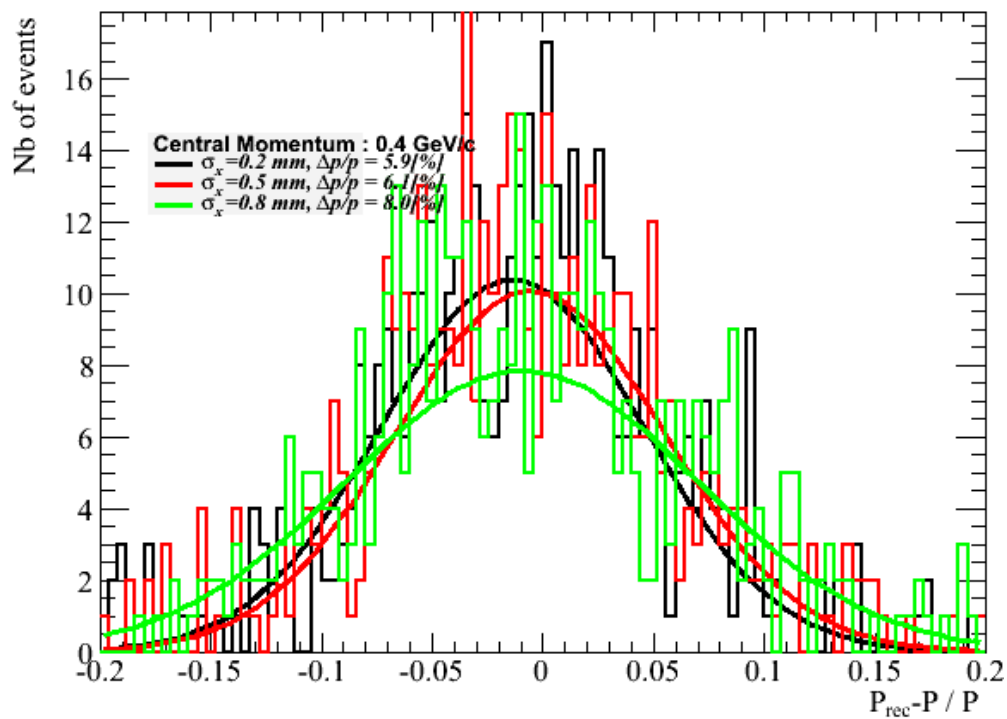

**Figure 20: Reconstructed momentum byte for a beam of 0.4 GeV, taking into account the material on the beam line.**

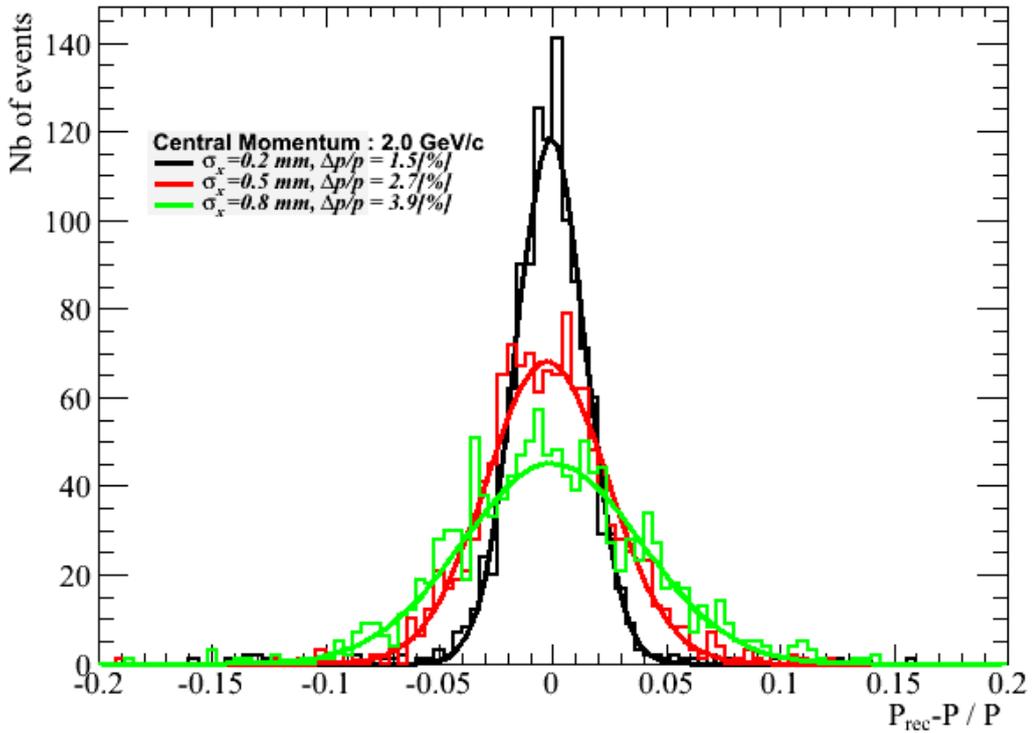

**Figure 21: Reconstructed momentum byte for a beam of 2 GeV, taking into account the material on the beam line.**

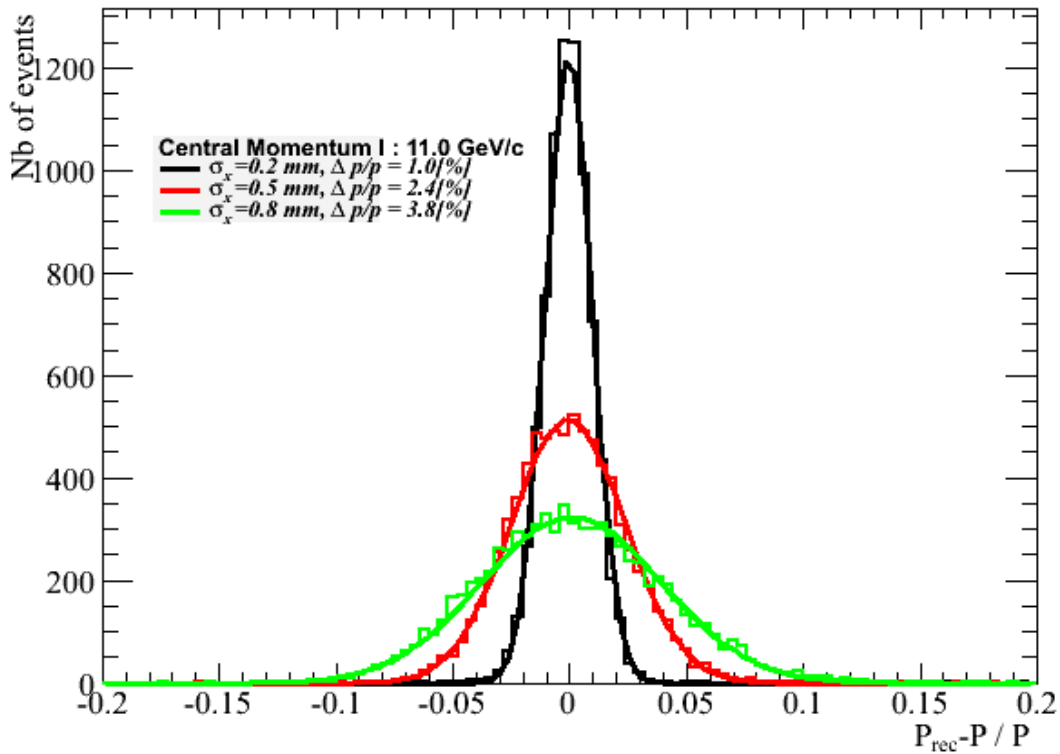

**Figure 22: Reconstructed momentum byte for a beam of 11 GeV, taking into account the material on the beam line.**

Figure 23 summarizes the results on the momentum reconstruction resolution over the full energy range of the beam and for three detector resolutions. For the expected resolution of the fiber profile detectors of 0.3 mm, the momentum can be determined to approx. 2% in the range from 1-12 GeV. We therefore conclude that the momentum measurement with the spectrometer can significantly improve the Δp/p of the beam without reducing the counting rate, and therefore

- 20 -

can be considered if precision measurements on the beam particles is required by the experiments.

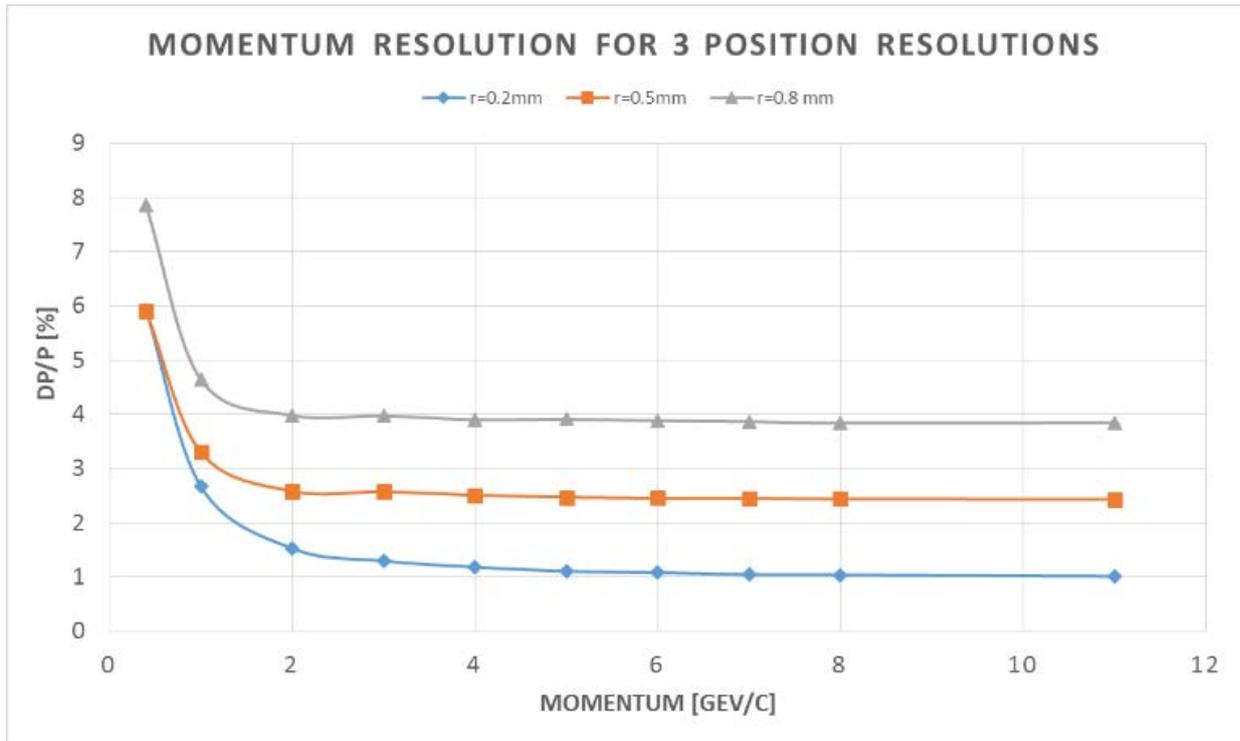

**Figure 23: Momentum resolution of the spectrometer for three different position resolutions, namely 0.2, 0.5 and 0.8 mm**

Although the beam particle momentum can be precisely determined as demonstrated above, the material in the line downstream the spectrometer will also deteriorate the particles' momentum reaching the active volume of the detectors. Figure 24 shows the ratio between the beam momentum in the middle of the cryostat (the liquid argon inside the detector itself is ignored) and the momentum at the position of the TOF0 detector. At low energies, a significant tail and shift in the mean momentum is observed that should be considered in the data analysis by the experiment, in particular for the low-energy electron beams.

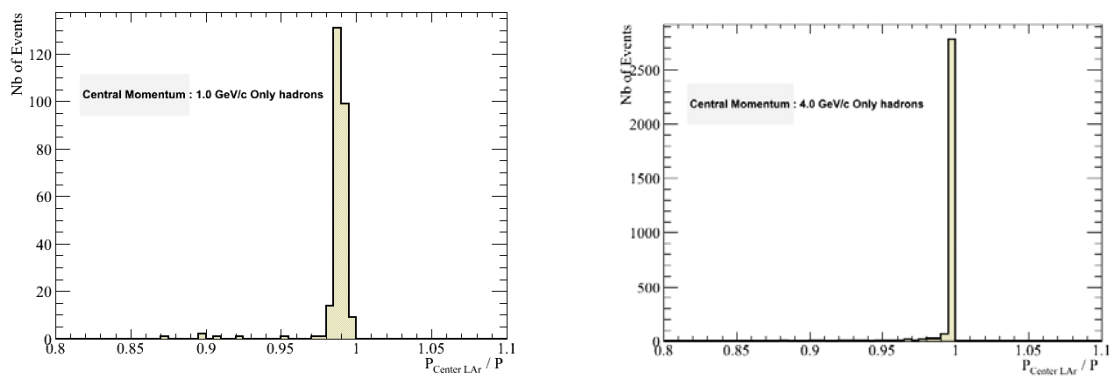

**Figure 24: Deterioration of particles' momenta due to the material present in the line, for the case of 1 and 4 GeV. The effect is more pronounced in the lower energies.**

## 4.2 Particle Identification (PID) with Cherenkov Counters and TOF

The proposed PID system for the VLE beam, uses a combination of two 2 m long threshold Cherenkov counters (XCET), and a TOF system with a flight path of 32 m, in order to tag all particles (in particular pions) cover he overall momentum spectrum provided by the beam line. The first Cherenkov XCET1 is filled with **Freon-12** and the second XCET2 with **$CO_2$**. They will operate in the range of pressures from ~100 mbar 10 bar, depending on the beam momentum and particle ID tagging requirement. The choice of Freon-12 was necessary due to the high refractive index providing satisfactory number of Cherenkov photoelectrons, in addition to its non-flammable properties.

In Figures 25 and 27 the pressure threshold as a function of the momentum and the particle type are shown for Freon-12 and $CO_2$ gases. For the XCET1 (Freon12) counter, pions emit radiation above 1.8 GeV for a pressure of 3 bar, kaons from 3.5 GeV for a pressure of ~ 10 bar and protons do not emit any radiation for pressures below 3 bar. For the XCET2 (CO2) counter, pions emit radiation above 3.5 GeV for a pressure of 2 bar, kaons from 4.5 GeV for a pressure of 15 bar and protons do not emit any radiation for pressures below 7 bar. In Figures 26 and 28 the number of photoelectrons for the interesting momentum regions are shown. For this calculation we assumed a 50% overall light collection and photoelectron production efficiency, to preserve a safety margin. It is clear that the Cherenkov system cannot tag pions and protons below 2GeV/c and a TOF system is needed.

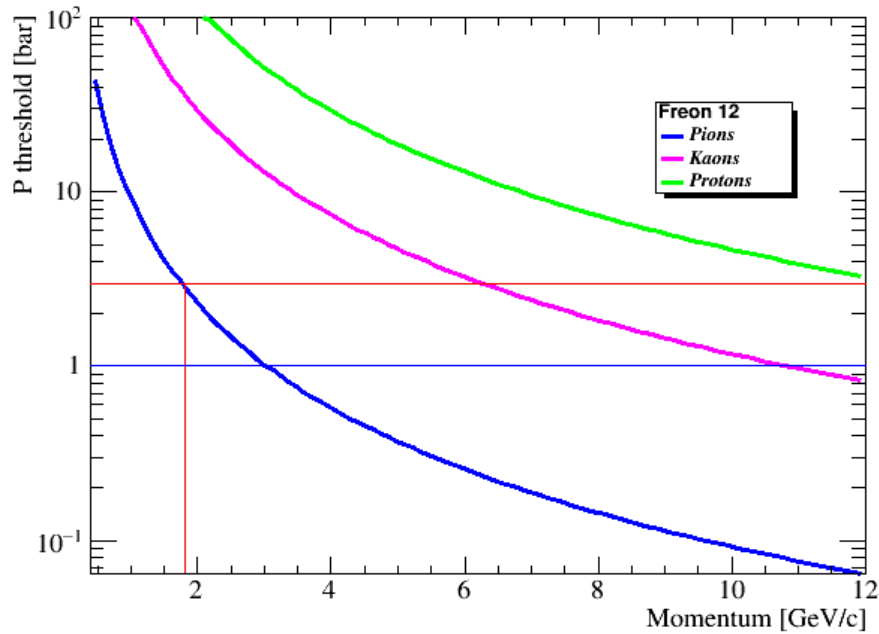

**Figure 25: Threshold pressure for the Cherenkov Counters filled with Freon 12 gas.**

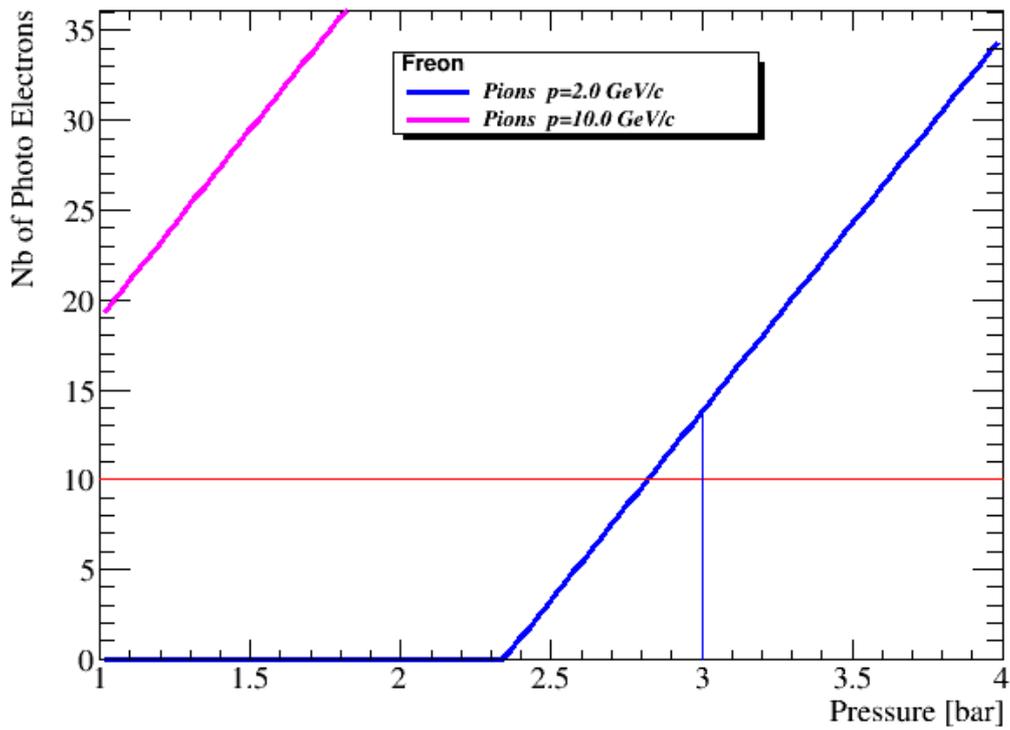

**Figure 26: Number of photoelectrons for pions for the XCET1 (Freon 12) Cherenkov.**

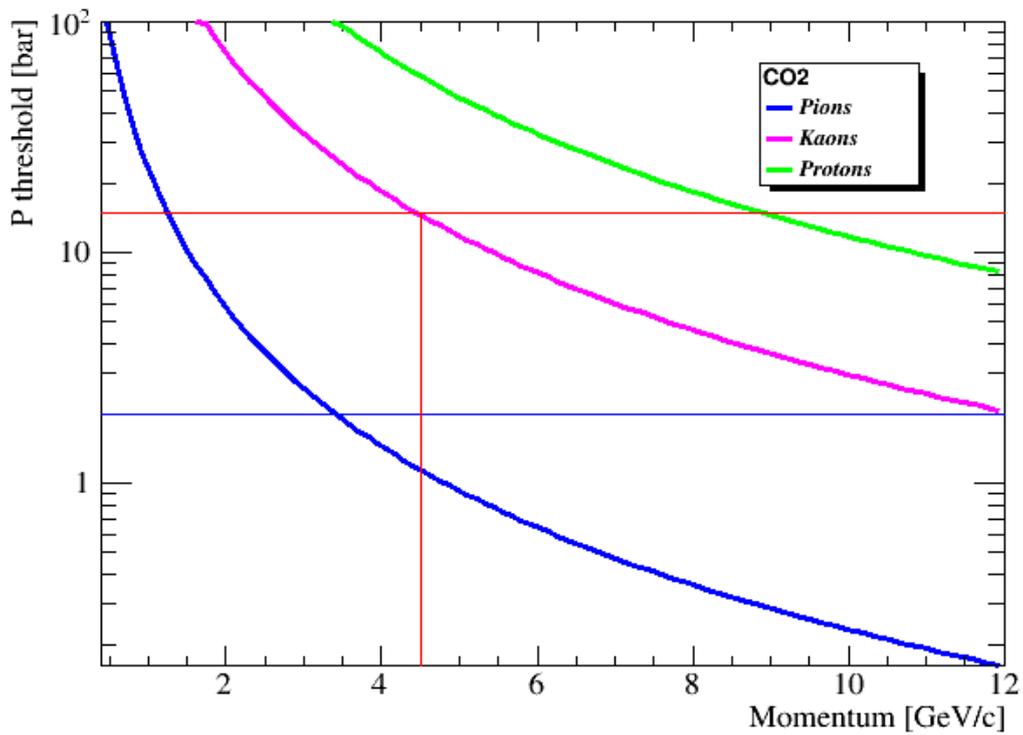

**Figure 27: Threshold pressure for the Cherenkov Counters filled with $CO_2$ gas.**

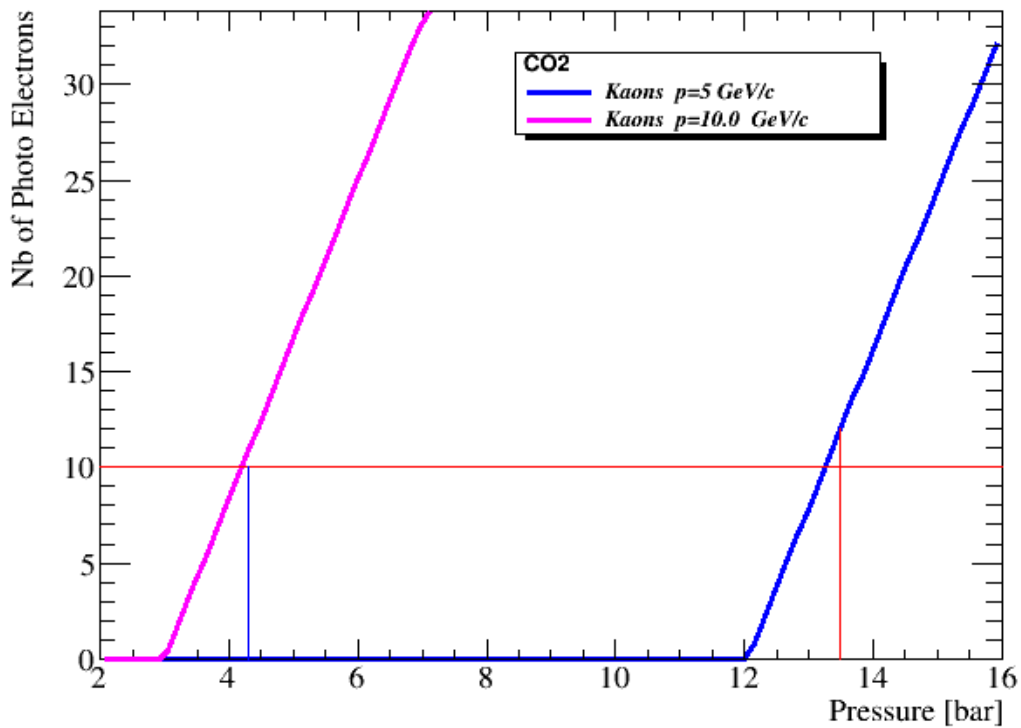

**Figure 28: Number of photoelectrons for kaons in the XCET2 ($CO_2$) Cherenkov counter**

In Figures 29 and 30, the TOF separation in terms of standard deviations, for pi-p and p-K separation are shown. The safe 4σ separation for pi-p tagging, is achieved with 2 ns TOF timing resolution for particles below 3 GeV, where 500 ps are needed for particles with momenta up to 5 GeV. For the p-K tagging, the 4 σ separation, is obtained with 500 ps TOF timing resolution for particles below 3 GeV, where 100 ps are needed for particles with momenta up to 6 GeV.

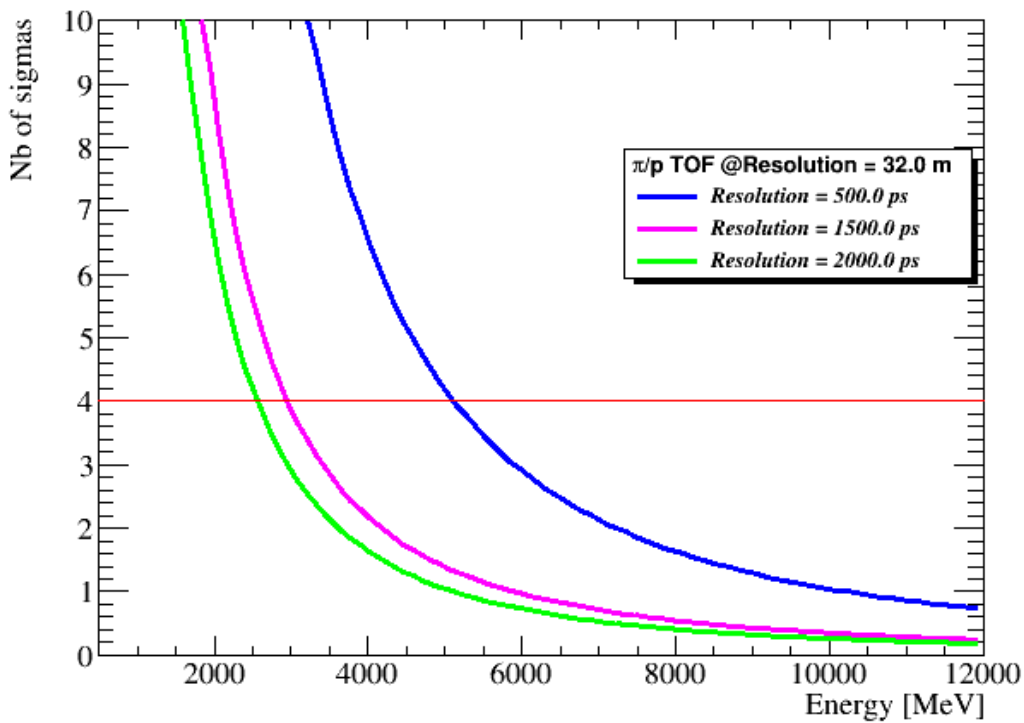

**Figure 29: Pion / proton separation using a time of flight system with the counters placed at 32 meters distance.**

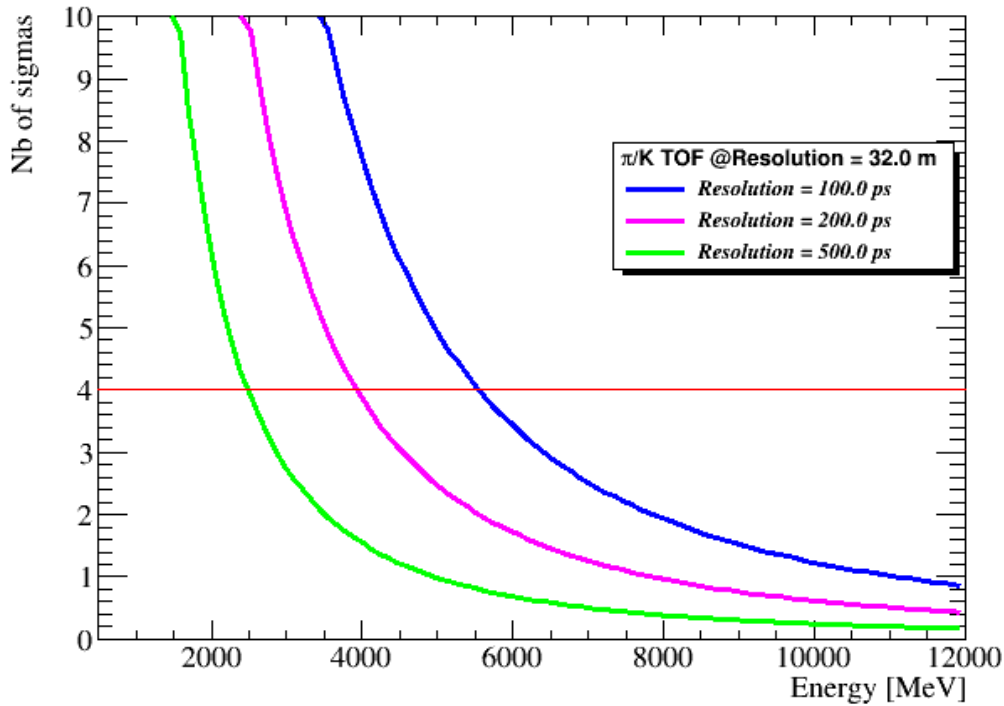

**Figure 30: Pion / kaon separation using a time of flight system with the counters placed at 32 meters distance.**

To summarize, the operational scenario of the combined PID system with the XCET/TOF detectors, we distinguish the following momentum regions:

**a. 0.4 – 2.0 GeV/c**

In this momentum region, we can only use the TOF system to identify pions from protons. Due to the short decay length of kaons in this low momentum range, the kaon population can be neglected. A relative low performance TOF is enough, and we aim for a timing resolution of ~1ns, as shown in Figure 29.

**b. 2.0 – 3.0 GeV/c**

In this momentum region, the first Cherenkov counter XCET1 is filled with Freon 12 at a pressure of 3 bars, while XCET2 remains empty. Pions with momenta in this region emit approximately 10 photoelectrons. Protons do not emit Cherenkov radiation at such a low pressure. Kaons below 3 GeV do not reach the experiment due their decay length, and therefore can be neglected.

**c. 3.0 – 4.5 GeV/c**

In this momentum region, XCET1 is filled with Freon 12 at 1 bar and XCET2 is empty. Pions above 3 GeV emit Cherenkov radiation, but protons and kaons do not. If we use a positive secondary beam to produce a positive VLE beam, as shown at Table 3, protons account for approx. ~7.7 % while kaons account for 0.64 – 1.46 %. Then the kaon contamination to protons, if no tagging is supplied varies from ~8 % to 16%. Alternatively, if we flip the sign of the VLE beam to negative, then kaon production, although low, dominates the anti-proton one, leading to a low contamination. An aerogel Cherenkov counter for an improved particle identification in this region is under investigation.

**d. 4.5 – 12.0 GeV/c**

In this highest momentum region, XCET1 is filled with Freon 12 at a pressure of 1 bar and XCET2 filled with $CO_2$ at a pressure of 13.5 bar, which will be gradually lowered as the momenta are increasing. In this configuration, pions emit radiation in XCET1 but not in XCET2, kaons emit in XCET2 but not in XCET1 and protons do not emit radiation in none of the detectors.

Electrons will always emit light. To enhance the data recorded with hadrons with low momenta, below 3 GeV, the XCET2 Cherenkov filled with $CO_2$ at 1 bar, maybe used as anti-coincidence in the trigger.

## 5. Summary

The beam particle composition, momentum measurement and particle identification methods to be used in the future H2-VLE (and similarly H4-VLE) beam line have been presented and discussed. The necessary beam instrumentation has been presented as well as our proposal for the PID detectors and their operation scenario. The momentum measurement can be done using the three profile monitors as spectrometer around the last bending dipole of the beam, for improved the momentum resolution, while the particle identification (tagging) using two threshold gas Cherenkov counters and a time of flight system can cover the full momentum range. The muon background as well as the details of the dedicated electron operation of the beam lines will be discussed in a following note.

## Acknowledgements

The authors would like to thank *Inaki Ortega-Ruiz* and *Jens Spanggaard* for providing details on the beam instrumentation and expected performance, and the colleagues in the EN-EA group for the numerous discussions. We wish also to thank the *LAPP computing department*, for their support allowing us to run thousands of CPU hours on the local TIER2 cluster.

## 7. Appendix

### 7.1 Cherenkov counters gases and pressures

| Momentum [GeV] | Target | XCET 1 | XCET2 | Density Freon [g/cm3] | Density CO2 [g/cm3] |
|---|---|---|---|---|---|
| 0.4 | W, R=30, L=300 | Vacuum | CO2 @ 1 bar | N/A | 0.001815 |
| 1.0 | W, R=30, L=300 | Vacuum | CO2 @ 1 bar | N/A | 0.001815 |
| 2.0 | W, R=30, L=300 | FR-12 @ 3bar | CO2 @ 1 bar | 0.014009 | 0.001815 |
| 3.0 | Cu, R=30, L=300 | FR-12@1bar | CO2 @ 1 bar | 0.004278 | 0.001815 |
| 4.0 | Cu, R=30, L=300 | FR-12@1bar | CO2 @ 1 bar | 0.004278 | 0.001815 |
| 5.0 | Cu, R=30, L=300 | FR-12@1bar | CO2 @ 13.5 bar | 0.004278 | 0.0266845 |
| 6.0 | Cu, R=30, L=300 | FR-12@1bar | CO2 @ 4 bar | 0.004278 | 0.00741075 |
| 7.0 | Cu, R=30, L=300 | FR-12@1bar | CO2 @ 4 bar | 0.004278 | 0.00741075 |
| 8.0 | Cu, R=30, L=300 | FR-12@1bar | CO2@ 4 bar | 0.004278 | 0.00741075 |
| 9.0 | Cu, R=30, L=300 | FR-12@1bar | CO2 @ 4 bar | 0.004278 | 0.00741075 |
| 10.0 | Cu, R=30, L=300 | FR-12@1bar | CO2 @ 4bar | 0.004278 | 0.00741075 |
| 11.0 | Cu, R=30, L=300 | FR-12@1bar | CO2 @ 2bar | 0.004278 | 0.00368025 |
| 12.0 | Cu, R=30, L=300 | FR-12@1bar | CO2 @ 2bar | 0.004278 | 0.00368025 |

http://www.peacesoftware.de/einigewerte/einigewerte_e.html - Temperature always 293.15K

### 7.2 Material Budget at 5GeV/c, and XCET2 at 15 bars pressure

| Beam Element | Nature | Size(cm) | Nb | Total Thickness (cm) | λ(%) | X0 (%) |
|---|---|---|---|---|---|---|
| After target | Air | 110 | 1 | 110 | 0,15% | 0,36% |
| S1 | Scintillator | 0,2 | 1 | 0,2 | 0,25% | 0,47% |
| TOF0 | Fibers | 0,1 | 1 | 0,1 | 0,13% | 0,23% |
| BPROF1 | Fibers | 0,1 | 1 | 0,1 | 0,13% | 0,23% |
| BPROF2 | Fibers | 0,1 | 1 | 0,1 | 0,13% | 0,23% |
| BPROF3 | Fibers | 0,1 | 1 | 0,1 | 0,13% | 0,23% |
| S2 | Scintillator | 0,2 | 1 | 0,2 | 0,25% | 0,47% |
| Before XCEt1 | Air | 3 | 1 | 3 | 0,00% | 0,01% |
| XCET1 | Al | 0,085 | 2 | 0,17 | 0,43% | 1,91% |
| XCET1 | Freon12 | 200 | 1 | 200 | 0,81% | 3,62% |
| Before XCET2 | Air | 3 | 1 | 3 | 0,00% | 0,01% |
| XCET2 | Al | 0,085 | 2 | 0,17 | 0,43% | 1,91% |
| XCET2 | CO2 at 15bars | 200 | 1 | 200 | 4,42% | 3,62% |
| Before BPROF4 | Air | 3 | 1 | 3 | 0,00% | 0,22% |
| BPROF4 | Fibers | 0,1 | 2 | 0,2 | 0,25% | 0,47% |
| S3 | Scintillator | 0,2 | 1 | 0,2 | 0,25% | 0,47% |
| Membrane | SS | 0,12 | 1 | 0,12 | 0,59% | 6,82% |
| **Total** | | | | | **6,98%** | **12,94%** |